\documentclass[11pt, a4paper]{article}
\usepackage[preprint]{tmlr}
\usepackage{parskip}
\usepackage[utf8]{inputenc}
\usepackage[T1]{fontenc}
\usepackage{graphicx}
\usepackage{eso-pic}  
\usepackage[margin=1in]{geometry}
\usepackage{amsmath}
\usepackage{amssymb}
\usepackage{hyperref}
\usepackage{transparent}

\usepackage{booktabs}
\usepackage{rotating}
\usepackage{array}
\usepackage{enumitem}  
\usepackage{todonotes}  
\usepackage{cleveref}  
\usepackage{tablefootnote}  
\usepackage{colortbl}

\usepackage{tcolorbox}
\usepackage{xcolor}

\usepackage{pdfpages}

\definecolor{custombeige}{HTML}{fff1ea}
\definecolor{customblue}{HTML}{d0e2f3}

\newcounter{example}[section]
\renewcommand{\theexample}{\thesection.\arabic{example}}

\newtcolorbox{exampleboxbeige}{
  colback=custombeige,
  colframe=black,
  boxrule=1pt,
  arc=0pt,
  left=10pt,
  right=10pt,
  top=10pt,
  bottom=10pt,
  before upper={\refstepcounter{example}\textbf{Example \theexample}\medskip}
}

\newtcolorbox{exampleboxblue}{
  colback=customblue,
  colframe=black,
  boxrule=1pt,
  arc=0pt,
  left=10pt,
  right=10pt,
  top=10pt,
  bottom=10pt,
  before upper={\refstepcounter{example}\textbf{Example \theexample}\medskip}
}

\renewcommand{\footnoterule}{%
  \vfill  
  \kern-3pt
  \hrule width 0.4\columnwidth
  \kern 2.6pt
}

\crefname{example}{Example}{Examples}
\Crefname{example}{Example}{Examples}
\crefname{table}{Table}{Tables}
\crefname{figure}{Figure}{Figures}  
\crefname{section}{Section}{Sections}
\crefname{equation}{Equation}{Equations}

\usepackage{etoolbox}
\AtBeginEnvironment{tabular}{\small}

\usepackage{footmisc}
\usepackage{titlesec}

\title{Risk Analysis Techniques for Governed LLM-based Multi-Agent Systems}
\author{Alistair Reid, Simon O'Callaghan, Liam Carroll and Tiberio Caetano}  
\date{29 July 2025}

\begin{document}

\begin{center}
\includegraphics[width=0.7\textwidth]{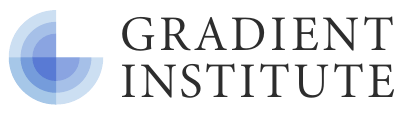}
\end{center}
\vspace{0.1\paperheight}
\AddToShipoutPictureBG*{
\AtPageUpperLeft{
\raisebox{-0.87\paperheight}{
\hspace*{-3mm}\includegraphics[width=\paperwidth]{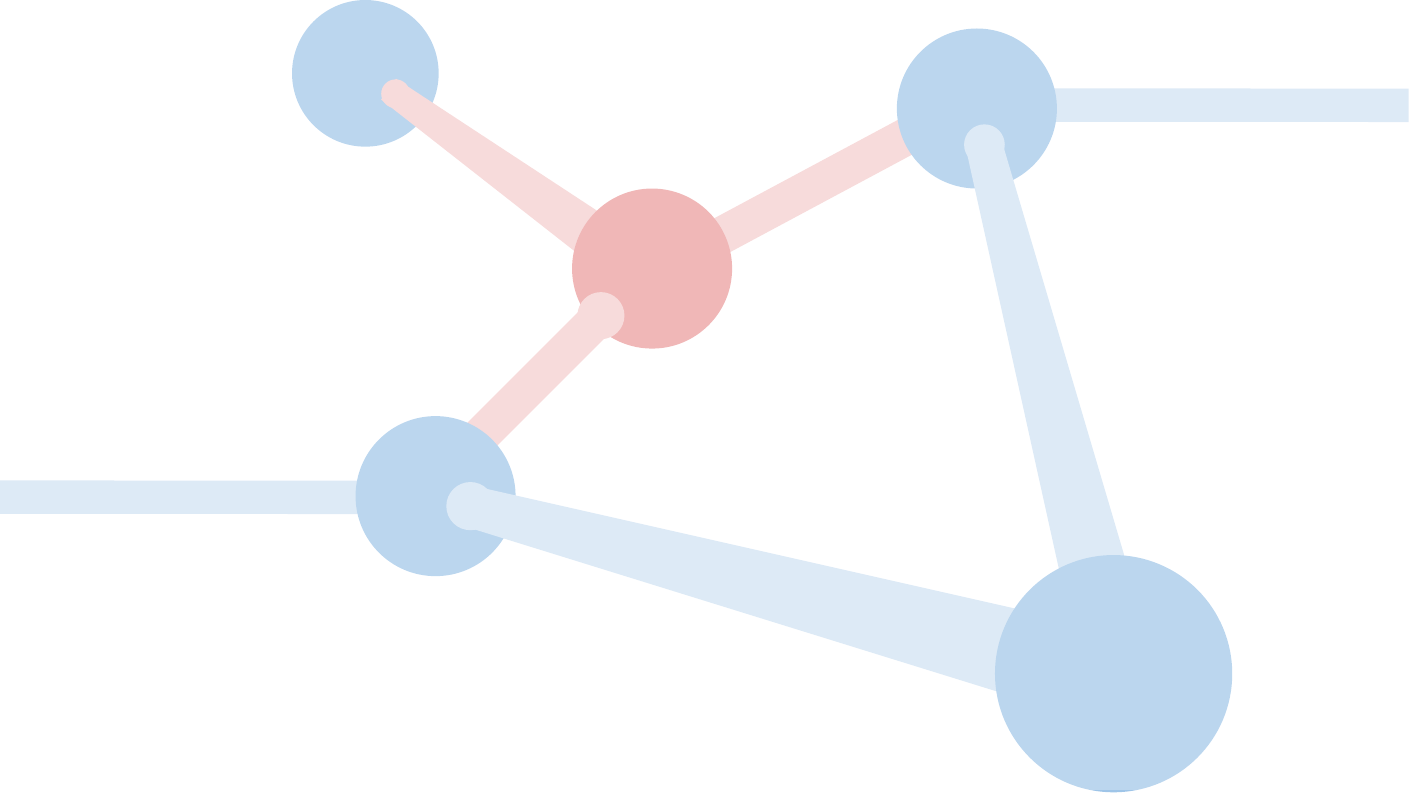}
}
}}
\makeatletter
\begin{center}
\vspace{0.3in}
{\fontsize{22}{30}\selectfont \parbox{0.7\textwidth}{\centering \textbf{\@title}}}
\\[3.65in]
{\fontsize{13}{20}\selectfont \textbf{\@author}}\\[3em]
{\fontsize{16}{22}\selectfont \textbf{\@date}}
\end{center}
\makeatother
\thispagestyle{empty} 
\newpage

\vspace*{1cm} 

{\centering
    \Huge \bfseries Risk Analysis Techniques for Governed LLM-based \\ Multi-Agent Systems \par
}

\vspace{1.0cm} 

Copyright © Gradient Institute Ltd. 2025 \\
Provided under a Creative Commons Attribution 4.0 International License. \\

This report has been prepared by Gradient Institute and supported by the Australian Government’s Department of Industry, Science and Resources. \\

\vspace{1.0cm} 
{\Large \textbf{Authors} \\}

Alistair Reid, Simon O'Callaghan, Liam Carroll and Tiberio Caetano
\vspace{1.0cm} 

{\Large \textbf{Acknowledgements}\\}

The authors would like to acknowledge Chris Dolman for regular discussions that helped shape the ideas reflected in this report, and Yaya Lu for assistance in figure preparation.\\

For detailed feedback on earlier versions of this report, the authors would like to thank: Ali Akbari, Bill Black, Alberto Chierici, Chris Dolman, Mark Hiscocks, Omid Karr, Willem Paling, Bronte Pendergast, Elija Perrier, Daniel Roelink, Jonathan Shen, Bill Simpson-Young, Kimberlee Weatherall, Kate White and Liming Zhu.
\vspace{1.0cm} 

{\Large \textbf{Contact}\\}

Gradient Institute Ltd. \\
\href{https://gradientinstitute.org}{https://gradientinstitute.org}\\
info@gradientinstitute.org
\clearpage

\section*{Executive Summary}
Organisations are starting to adopt AI agents based on large language models to automate complex tasks, with deployments evolving from single agents towards multi-agent systems. While this promises efficiency gains, \textbf{multi-agent systems fundamentally transform the risk landscape} rather than simply adding to it. \textbf{A collection of safe agents does not guarantee a safe collection of agents} – interactions between multiple LLM agents create emergent behaviours and failure modes extending beyond individual components.

This report provides guidance for organisations assessing the risks of multi-agent AI systems operating under a\textbf{ governed environment}, such that there is control over the configuration and deployment of all agents involved. We focus on the critical early stages of risk management – \textbf{risk identification and analysis} – offering tools practitioners can adapt to their contexts rather than prescriptive frameworks.

Six key failure modes emerge as particularly salient in governed multi-agent environments.\textbf{ Cascading reliability failures} manifest when agents' erratic competence and brittle generalisation failures are propagated and reinforced across the network.\textbf{ Inter-agent communication failures} involve misinterpretation, information loss, or conversational loops that derail task completion. \textbf{Monoculture collapse} emerges when agents built on similar models exhibit correlated vulnerabilities to the same inputs or scenarios. \textbf{Conformity bias} drives agents to reinforce each other's errors rather than providing independent evaluation, creating dangerous false consensus. \textbf{Deficient theory of mind} occurs when agents fail to incorporate correct assumptions about other agent's knowledge, goals or behaviours, leading to coordination breakdowns. \textbf{Mixed motive dynamics} arise when agents pursuing individually rational objectives produce collectively suboptimal outcomes, even under unified governance.

Our approach to analysing these failure modes centres on the concept of \textbf{validity} – the multifaceted quality encompassing whether assessments measure what they claim to measure, produce consistent and reliable results, align with real-world outcomes, and yield insights that are both theoretically sound and practically meaningful. Given fundamental limitations in the scientific understanding of LLM behaviour, practitioners must work towards \textbf{progressively increasing validity through convergent evidence} across multiple assessment approaches. This involves \textbf{staged testing that gradually increases exposure} to potential negative impacts, combining simulations with observational data analysis, benchmarking against appropriate baselines, and red teaming to uncover hidden vulnerabilities.

This report equips organisations with foundational knowledge and tools for analysing the distinctive challenges that emerge when multiple AI agents work together within their governed environments. While translating from risk analysis to risk evaluation and treatment remains highly  contextual and organisation-specific, these methodologies provide a starting point for developing robust multi-agent risk management practices as these systems become increasingly prevalent across critical business functions.

\newpage
\tableofcontents
\thispagestyle{empty} 

\newpage
\setcounter{page}{1} 


\section{Introduction}\label{sec:introduction}

\subsection{Context of this Report}

{\large \textbf{\textit{Multi-agent AI systems under organisational governance}}}

Large language models (LLMs) are rapidly evolving from conversational applications to autonomous LLM agents capable of executing complex tasks with minimal human supervision. Organisations are starting to deploy LLM agents to automate complex tasks, as they have the potential to provide efficient and low-cost automation of business processes, to free professionals to focus on high-value activities and to provide sophisticated collaborative assistance for complex problem-solving.

LLM agents are systems that combine large language models with software scaffolding (such as tools, communication protocols and memory systems for context retention) for autonomous task execution. They represent a significant shift from traditional automation in terms of flexibility and adaptability, but often at the cost of predictability and interpretability.

Furthermore, single-agent deployments are evolving into multi-agent deployments. Organisations have compelling reasons to deploy multiple, interacting, specialised agents. Different business units naturally automate their domains independently, and can incrementally replace existing processes with LLM agents. Those business units do, however, interact: an HR agent processing new hires must coordinate with IT agents provisioning access and finance agents setting up payroll. Such cross-functional dependencies suggest that enabling direct agent-to-agent communication could streamline operations beyond what isolated agents could achieve. Even within single departments this can still hold – customer service might deploy separate agents for initial triage, technical troubleshooting, and escalation handling, each specialising in distinct aspects of the support workflow while needing to share context and coordinate handoffs.

Within organisational boundaries, these multi-agent systems can operate under common governance – a shared framework of oversight, coordination protocols, and aligned objectives determining how different agents are configured and deployed. This distinguishes them from multi-agent scenarios where interacting agents can be deployed by different organisations, individuals, or unaffiliated groups, potentially lacking shared communication protocols, operating without coordinated oversight mechanisms, or pursuing misaligned or adversarial objectives. 

\textbf{In this report we focus on multi-agent systems in the context of a single organisation, and assume the organisation has control over how the agents are configured and deployed} – we assume a governed setting for multi-agent AI systems. 

\pagebreak

\subsection{Premise of this Report}

{\large \textbf{\textit{Multi-agent AI systems introduce distinctive risks}}}

Even when we assume a multi-agent system is operating under common governance, distinctive risks arise. Importantly, \textbf{a collection of safe agents does not imply a safe collection of agents.} The interaction between multiple agents creates emergent behaviours and failure modes that extend beyond those of individual agents. Even with unified oversight and aligned objectives, multi-agent systems can exhibit cascading errors, coordination failures, and unintended collective behaviours.

The multi-agent paradigm challenges traditional AI governance methods through several distinctive characteristics:
\begin{itemize}
    \item \textbf{Emergent behaviours and novel failure modes:} Multi-agent systems generate entirely new categories of failure through group dynamics that cannot exist in single-agent deployments.
    \item \textbf{Amplification of existing risks: }Multi-agent systems don't just accumulate single-agent risks; they transform them. Cognitive limitations become cascading hallucinations, communication errors become system-wide miscoordination, and individual biases become collective blind spots through reinforcement dynamics.
    \item \textbf{Shifting control paradigms:} Humans may increasingly delegate decisions to networks of autonomous agents, reducing opportunities for direct oversight and creating potential governance gaps between agent interactions.
    \item \textbf{Limited precedent:} Governance of LLM-based multi-agent systems is a nascent field with emerging principles but no established standards or mature practices. Staff with responsibility for risk must navigate this uncertainty while making deployment decisions that could have significant organisational impact.
\end{itemize}

These characteristics mean that analysing single agents cannot ensure multi-agent system safety. Multi-agent environments demand governance frameworks specifically designed to manage emergent behaviours, prevent risk amplification, and maintain meaningful oversight across networks of interacting agents. 

\pagebreak
\subsection{Purpose of this Report}
{\large \textbf{\textit{Guidance on risk analysis for governed multi-agent AI systems} 
}}

Organisations face an important challenge: how to integrate AI risk into their risk management frameworks? The purpose of this report is to support organisations in addressing this challenge by providing guidance on AI risk identification and analysis – specifically for multi-agent AI systems within the governed environment of a single organisation. 

This report is intended to benefit a range of organisational stakeholders, from system owners and those responsible for AI system actions, to organisational leadership including board-level executives, and risk management practitioners who need practical guidance for incorporating AI considerations into their existing risk frameworks.

Our focus is on the early stages of the risk management pipeline – specifically risk identification and risk analysis. The later stages comprising risk evaluation and treatment require deeper contextual understanding of a specific deployment and are outside the scope of this report, which aims to be practical and yet generally applicable. In the discussion we will elaborate on considerations for integrating the tools we offer in this report into the later stages of the risk management pipeline.

We specifically examine LLM-based multi-agent systems operating in governed environments where agents cooperate or carry out complementary roles under shared organisational or technical oversight. We introduce the emerging landscape of multi-agent systems, present key information for identifying potential failure modes, and provide methodological tools and practices for analysing these risks. 

Through this report, practitioners will gain awareness of failure modes specific to multi-agent interactions, learn general principles and techniques for assessment of these failure modes, understand current measurement limitations and capabilities, and acquire tools to update their risk assessment frameworks accordingly. 

\textbf{Our aim is to equip organisations with foundational knowledge for navigating the distinctive challenges that arise when multiple AI agents work together within their operational boundaries.}

\pagebreak
\subsection{Scope and Approach of this Report} \label{sec:intro-scope}

Several key decisions shape both the scope and approach of our risk analysis:

\textbf{Focus on governed environments: }We consider agents operating under common organisational governance – where they cooperate or carry out complementary roles within a shared framework of oversight and coordination. This excludes adversarial scenarios or fully open multi-agent markets where agents lack unified governance. We acknowledge that focusing on governed systems represents an initial step in understanding the broader multi-agent risk landscape. Future work should address separately governed scenarios where coordination challenges multiply and new strategic risks emerge – from business partnerships to competitive markets, where agents pursue potentially conflicting objectives. However, analysing governed systems provides the essential foundation for this broader analysis.

\textbf{Agent-to-agent dynamics:} Our analysis centres on interactions between agents rather than human-agent collaboration. While real deployments will include human participants as supervisors, team members, or decision points – factors that can affect the risk profiles – we focus primarily on the agent-to-agent dynamics that create distinctive multi-agent risks. Many of our analysis techniques remain applicable to hybrid human-agent systems.

\textbf{Emphasis on capability limitations over potential for deception: }This report predominately emphasises risks arising from model unreliability - such as cognitive limitations, communication failures or coordination breakdowns - rather than risks posed by sophisticated deceptive capabilities. This focus addresses the practical reality that, at present, most organisational risks stem from insufficient capabilities rather than overly capable manipulation or strategic deception against human interests. However, as models become increasingly capable, practitioners should remain aware of developments in the science of strategic deception risks, including recently published frameworks and evidence of AI scheming \citep{balesni_towards_2024, meinke_frontier_2025}.

\textbf{Identification and analysis, not evaluation:} This report focuses on the initial stages of risk management: identifying potential failure modes and \textbf{analysing} their likelihood of occurrence.\footnote{The term ‘evaluation’ carries distinct meanings in the AI and risk management communities.
\begin{itemize}
    \item In risk management, as defined by standards like ISO 31000 \citep{isotc_31000_2018}, risk evaluation is the decision-making phase that follows risk identification and analysis. Its central question is, ``Is this risk acceptable or does it need treatment?'', requiring deep contextualisation and judgment based on organisational risk appetite.
    \item In the AI community, model evaluation typically refers to the technical assessment of a system's performance and capabilities, often against narrow benchmarks and quantitative metrics that aim to assess competence in areas such as coding, math, or reasoning.
\end{itemize}
 }
The scope of our analysis does not extend to evaluating the potential impact or severity of these failures, nor does it propose treatments. Impact assessment is omitted because it is highly dependent on contextual factors (such as specific deployments, use cases, and organisational risk appetites) that are beyond the purview of this work.

Table \ref{tab:risk_phases} clarifies this distinction using ISO 31000 terminology \citep{isotc_31000_2018}.

\begin{table}[ht]
    \centering
    \caption{The risk assessment phases as outlined by ISO 31000 \citep{isotc_31000_2018}. This report contributes to identification and analysis of risk in a governed multi-agent setting.}
    \label{tab:risk_phases}
    \begin{tabular}{|p{0.2\textwidth}|p{0.2\textwidth}|p{0.25\textwidth}|p{0.25\textwidth}|}
        \hline
        \rowcolor{customblue}
        \textbf{Risk Phase} & \textbf{Focus} & \textbf{Key Question} & \textbf{Output} \\
        \hline
        \textbf{Identification} & Discovery & ``What could happen?'' & Set of salient failure modes \\
        \hline
        \textbf{Analysis} & Understanding & ``What can be measured about the impact and likelihood?'' & Measurements and methodologies pertaining to risks \\
        \hline
        \textbf{Evaluation} (out of scope as it is highly contextual to the use-case) & Decision-making & ``Is this risk acceptable or does it need treatment?'' & Prioritised risks and judgement on their acceptability and how to manage them. \\
        \hline
    \end{tabular}
\end{table}

\textbf{A toolkit, not a prescription: }We present techniques and practices for analysing key multi-agent failure modes, not an exhaustive framework or rigid methodology. Multi-agent risk assessment remains a nascent field with emerging principles but no mature standards. Practitioners must adapt these tools to their contexts and map identified failure modes to organisational impacts – the critical step for moving from risk analysis to risk evaluation.

\pagebreak
In this report, we distinguish between failure modes and risks in line with standard risk management principles. A failure mode is the specific mechanism through which a system can fail. It answers the risk identification question: ``What could happen?''. Identifying this set of potential failures is a primary step in the risk management process. A risk is a broader concept that contextualises a failure mode by considering its probability and the severity of its consequences. After a failure mode is identified, the subsequent risk analysis phase seeks to understand the associated risk by asking, ``What can be measured about the impact and likelihood?''
\\

\textbf{Note on anthropomorphic language} 

For certain concepts presented in this report, we follow the established practice in the AI research and practitioner community of using anthropomorphic framing to help map familiar organisational and cognitive concepts to AI agent behaviours. This widely-adopted approach helps create useful mental models of the underlying computational processes.  We may refer to agents ``planning,'' ``reasoning,'' ``coordinating,'' or ``pursuing objectives.'' This language choice leverages our pre-existing human understanding of analogous concepts. For instance, when we say agents ``plan,'' we mean the model produces outputs that, if produced by humans, would be associated with planning. \textbf{Crucially, however, we do not intend to imply that the underlying mechanisms are the same or that these systems possess human-like cognition.}

\pagebreak
\subsection{Structure of this Report}
This report builds understanding through four progressive sections:

\textbf{\cref{sec:foundations}: Foundations and Frameworks of LLM-Based Multi-Agent Systems} introduces LLM agents and presents canonical multi-agent settings – archetypal configurations that help practitioners identify which failure modes are most relevant to their systems. These settings provide a structured way to map real-world deployments to their associated failure modes.

\textbf{\cref{sec:failuremodes}: Failure Modes in Governed Multi-Agent Systems }details specific failure modes that emerge from agent interactions. We explain each failure mode's mechanisms, link them to the canonical settings where they're most likely to manifest, and provide examples of their potential real-world manifestations.

\textbf{\cref{sec:riskanalysis}: Analysis Techniques} provides practical tools and approaches for assessing the identified failure modes. We emphasise progressive testing stages – from simulation through deployment – while acknowledging the fundamental limitations of pre-deployment assessment. We also emphasise and elaborate on the key notion of validity, which provides a lens through which practitioners can assess the limits of a certain method or tool.

\textbf{\cref{sec:discussion}: Discussion} reflects on the broader implications of our analysis, including the relationship between multi-agent and human management, the limitations of our framework, its relevance to concerns about catastrophic risks, and considerations for practitioners working to integrate these insights into their organisational contexts.

Through this progression, we provide practitioners with both conceptual understanding and practical tools for analysing the distinctive risks that emerge when AI agents work together within governed environments.

\section{Foundations and Frameworks of LLM-Based Multi-Agent Systems}\label{sec:foundations}

This section introduces LLM agents and canonical multi-agent configurations, providing the foundational platform for understanding the failure modes of multi-agent systems.

\subsection{LLM Agents}

An \textbf{\textit{AI Agent}} is an AI system that interacts with an environment through observation and actions, often with the objective of achieving a specified goal. 

An AI agent is an autonomous system that is distinguished by its AI-driven adaptive decision making engine, rather than following a set plan. This means that AI agents are able to handle greater variability, but often at the cost of reduced reliability and interpretability. 

An important concept is the \textit{level of autonomy} of the agent \citep{feng_levels_2025} – the extent to which an AI agent is designed to operate without user involvement – which shapes its capability and risk profile. In addition to capturing the extent of user involvement, the broader notion of autonomy reflects how independently a system can operate. This includes the ability to observe its environment, make decisions, execute actions, and adapt behaviour with minimal human intervention or oversight.

Some examples of AI agents, in increasing order of complexity and autonomy, include:

\begin{itemize}
    \item \textbf{Game-playing agents }that evaluate situations and select strategic moves in board games and video games
    \item \textbf{Voice-activated assistants} such as Alexa, Siri and Google Assistant, that process natural language to perform tasks like playing music, sending texts, and controlling smart home devices
    \item \textbf{Self-driving vehicles} making real-time navigation and safety decisions on the road
    \item \textbf{Computer-using research assistants }that design and conduct experiments with minimal supervision by interacting with the web, executing code, and performing other computational tasks.
\end{itemize}

Historically, AI agents have been developed for a specific (narrow) application – chess engines cannot drive cars, and robotic controllers cannot write code.\textbf{ LLM agents }represent a fundamental departure by combining the general purpose and domain-agnostic capabilities of large language models (LLMs)\footnote{A large language model is an artificial neural network trained on vast quantities of text to predict the next word (or token) in a sequence. Through next-token prediction, LLMs exhibit a broad spectrum of knowledge and capabilities, which allows them to generate coherent and contextually relevant text to a given prompt. Some LLMs are multi-modal and able to process inputs such as images or audio as well as text.} with software scaffolding that enables autonomous environmental interaction, thereby empowering AI agents to tackle an unprecedented breadth of applications.

For instance, the same LLM agent framework might be applied (with an appropriate set of software tools) to\footnote{For example frameworks and applications, see \cite{huizenga_agent_2025}.}:

\begin{itemize}
    \item \textbf{Enterprise workflow automation} such as processing invoices or scheduling
    \item \textbf{Customer service systems} that can resolve inquiries using knowledge bases
    \item \textbf{Code development environments} to write, debug and execute code
    \item \textbf{Content management systems }to create and optimise content.
\end{itemize}

\begin{figure}[t]
    \centering
    \includegraphics[width=0.75\textwidth]{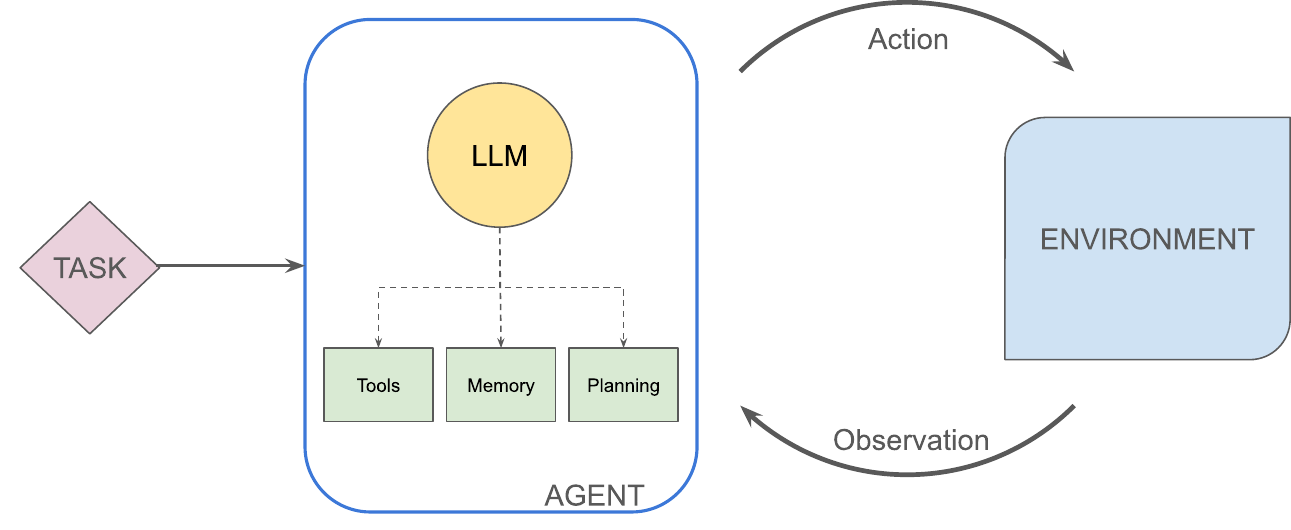}
    \caption{System diagram of an LLM Agent}
    \label{fig:agent_diagram}
\end{figure}

\textbf{LLM Agent Architecture: }A large language model serves as the cognitive engine of an LLM agent. Because LLMs can interpret complex instructions, reason through multi-step problems, and adapt their strategies based on natural language feedback, LLM-based agents are far more flexible and capable than traditional automated systems.

The scaffolding around an LLM agent typically includes several components:

\begin{itemize}
    \item \textbf{Action execution:} Software that translates LLM outputs into environmental actions
    \item \textbf{Environmental perception:} Systems that extract observations and feed them back into the LLM (for example, taking screenshots of a computer)
    \item \textbf{Tool interfaces:} Access to external software tools and APIs for task execution \citep{qu_tool_2025}
    \item \textbf{Memory and planning modules}: Enhanced context management (though often this simply involves how information is processed within the model's context window)
    \item \textbf{Reasoning frameworks: }Many LLM agents are prompted to follow iterative planning and execution frameworks like ReAct (Reason-Act-Observe) \citep{yao_react_2023}, which guides them through cycles of generating plans for their situation, taking actions, and observing results.
\end{itemize}

Additionally, if an LLM can handle multimodal inputs, then their applications expand beyond text-based environments to uses such as operating full computer systems, referred to as Computer-Using Agents \citep{openai_computer-using_2025}. 

While any LLM can theoretically be turned into an agent with appropriate scaffolding, the framework alone doesn’t guarantee competency as an agent. LLM agents often require additional training (such as a reinforcement learning phase) to effectively operate in their environment, use the software tools at their disposal, and form pathways to achieve specific goals. However, once properly developed, these systems enable a wide range of practical applications.

A system with LLM components is not necessarily an agent system as we have defined agents. For example,\textbf{ LLM workflows }are predefined execution pipelines that chain together LLM tasks that call APIs, invoke software tools, or execute scripts (see the figure on the next page). However, workflows do exhibit some analogous failure modes, such as error cascades.

LLM agents are particularly challenging subjects for risk management. Unlike traditional rule-based software, or even basic AI systems with more predictable failure modes, LLM agents combine the unpredictability of LLMs with the compounding effects of autonomous decision-making in real environments. These complexities become even more pronounced when multiple LLM agents interact and the decision making becomes decentralised, as discussed in \cref{sec:canonical_settings}.

\begin{figure}[p]
    \centering
    \includegraphics[width=0.9\textwidth]{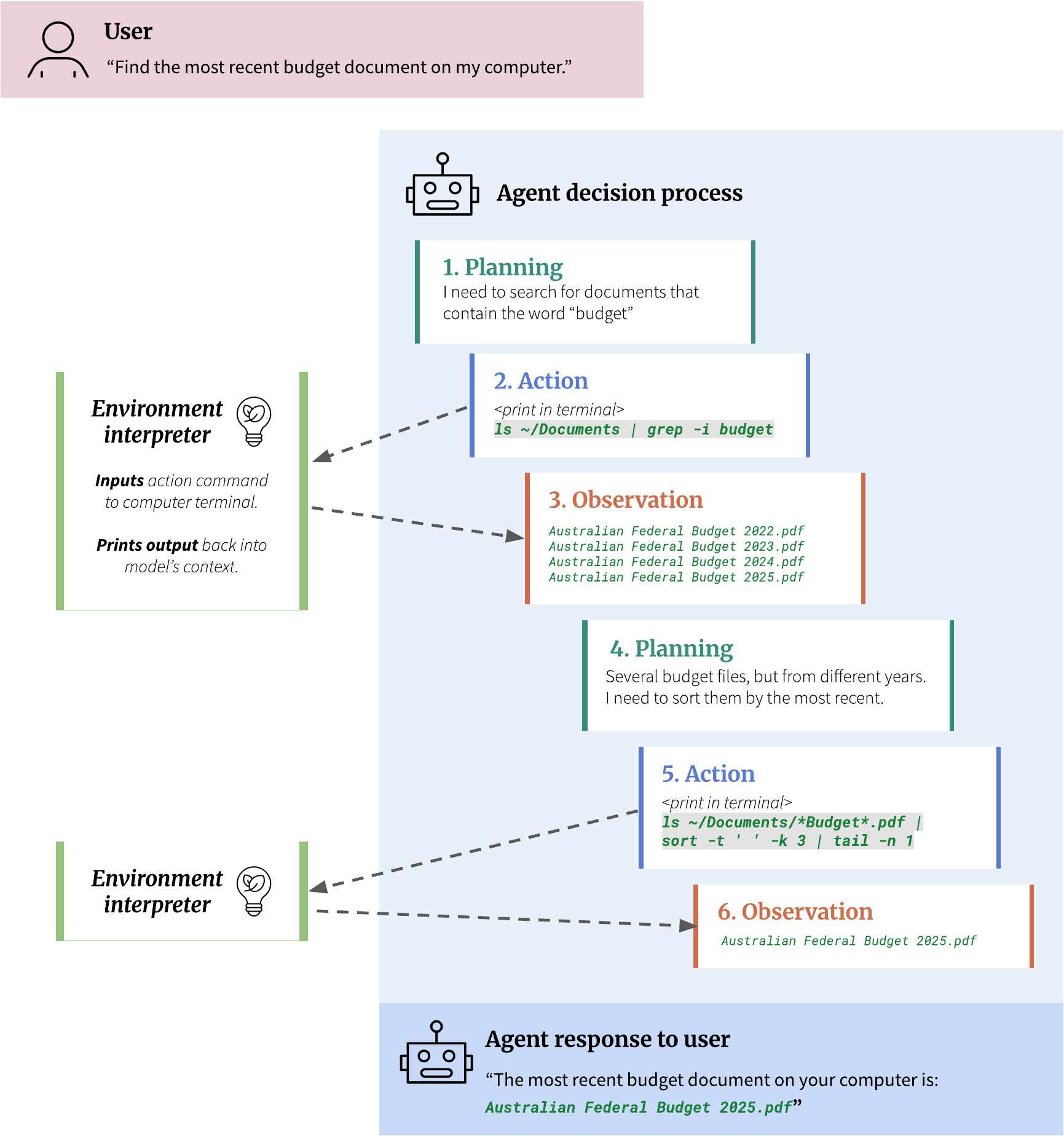}
    \setlength{\abovecaptionskip}{6pt}
    \caption{Example LLM agent workflow using the ReAct framework to plan, take action, then observe, in a loop
}
    \label{fig:agent_workflow}
\end{figure}

\newpage
\subsection{Canonical Multi-Agent Settings} \label{sec:canonical_settings}
In this section, we introduce a (non-exhaustive) selection of canonical multi-agent settings - reference cases for studying multi-agent system failure modes.
As introduced in \cref{sec:intro-scope}, we focus specifically on agents operating in a governed ecosystem under a shared organisational or technical framework. We assume that:

\begin{itemize}
    \item Agents are \textbf{designed to coexist and interact} through common infrastructure such as communication protocols
    \item Agents are subject to \textbf{shared governance} such as access control and resource allocation rules
    \item Agents \textbf{operate within a trusted network} of authenticated participants (where they may be known or unknown to the agent, if deployed by a different business unit for example).

\end{itemize}

The multi-agent settings featured in this section are simplified scenarios designed to help practitioners identify which failure modes they are most likely to encounter in their real systems. These scenarios distil complex real-world systems into archetypal patterns\footnote{Some elements of the system specification (such as the choice of LLM model) are critical to the behaviour of the system, but are excluded from this list because they do not alter which canonical setting the system maps to.} \footnote{ In practice, these settings may include human participants integrated as supervisors, domain experts, or collaborative partners alongside the AI agents. While human involvement can alter specific coordination dynamics and risk profiles, many of the failure modes and assessment approaches discussed in this report remain applicable to these hybrid systems of humans and agents.}.  The key distinguishing features between settings include:

\textbf{Network Topology - }The structure of connections between agents: who can communicate with whom. This could be a centralised layout (where all agents report to a hub), a fully connected network (where all agents talk to each other), or something modular (e.g. clusters or sub-teams).

\textbf{Objective Structure - }How goals and tasks are distributed across agents. Are all agents working towards the same shared objective, or are tasks divided into separate but related sub-goals with independent objectives? Tasks may be pre-assigned statically, allocated dynamically by a supervising agent, or selected by agents themselves.

\textbf{Communication Protocol - }How and how often agents exchange information. This includes the frequency of communication (continuous vs. event-based), the format (structured or unstructured), and whether communication is one-to-one, one-to-many, or broadcast.

\textbf{Agent Roles and Specialisations -} What each agent is capable of doing. Are agents interchangeable, or do they have specialisations (e.g. one writes text, another queries databases)? Do they have different access to software tools, knowledge, or data?

\textbf{Persistence and Memory -} Whether agents are short-lived or long-running. Do they remember past interactions and build up context over time, or do they start fresh each time they're used?

These considerations allow us to identify different archetypes of multi-agent systems in a structured way. Most real-world applications will resemble one or more of the canonical settings described below, which we then map to their most salient failure modes in \cref{sec:map-canonical-failure}.

\textbf{Knowing which setting a system most closely matches can help identify which kinds of failure modes warrant the most attention, and which types of risk analysis are most applicable.}

\pagebreak
\subsubsection{Single Agent Equivalent}

\begin{figure}[h!]
    \centering
    \includegraphics[width=0.65\textwidth]{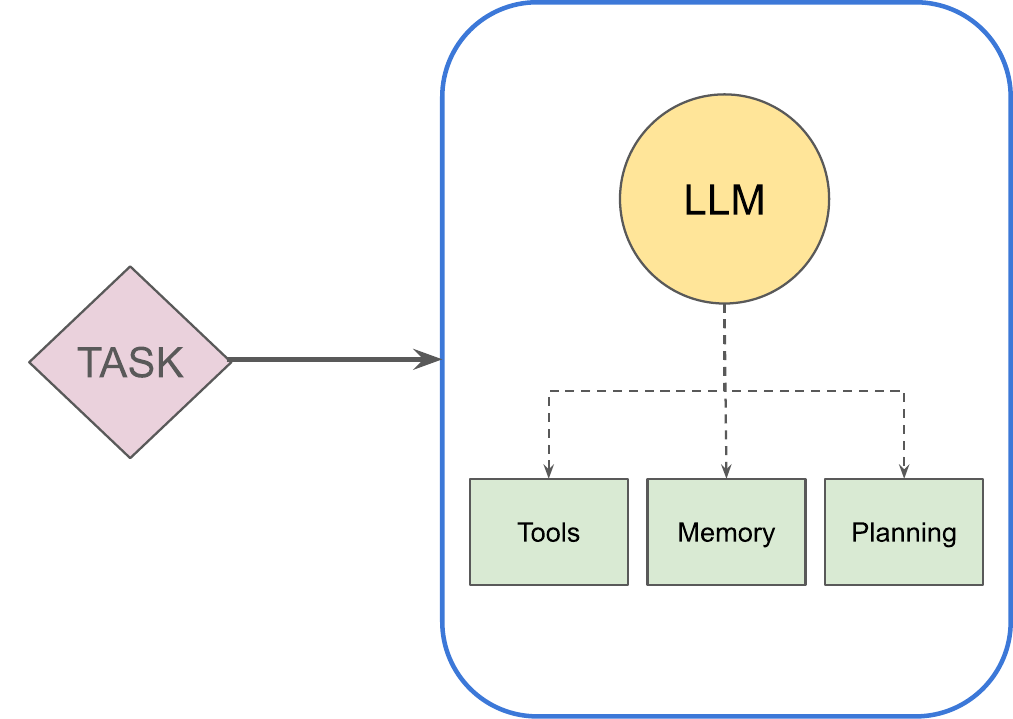}
    \caption{System diagram of a Single Agent Equivalent setting
}
    \label{fig:single_agent}
\end{figure}

This is the foundational setting: a single agent, acting autonomously, carries out a task end-to-end. It will interpret its instructions, and may use software tools and interact with external systems, but all control and planning output is localised within a single decision-making unit.

\textbf{Examples:}

\begin{itemize}
    \item Data retrieval, where a single agent queries software tools and synthesises answers
    \item Travel booking assistants that plan and complete bookings by interacting with APIs
    \item Email triage agents that categorise messages and draft replies
    \item Coding assistants that generate, explain, execute and debug code in response to developer input
\end{itemize}

Single-agent settings deserve careful attention because:

\begin{itemize}
    \item \textbf{Single-agent systems are not trivial.} They carry meaningful risks related to decision quality, tool use, misinterpretation of context, and task planning errors.
    \item \textbf{These risks carry forward. }Every multi-agent system is exposed to the failure modes of single-agents, as well as additional layers of cascading or emergent behaviour that arise from their interactions\footnote{ While this report focuses on how multi-agent systems can amplify risks, multi-agent configurations may sometimes help in mitigating single-agent failures. For instance, multiple agents voting on decisions or averaging their outputs can correct for individual errors, similar to ensemble methods in machine learning.}.
\end{itemize}

In short, considering single-agent failure modes is an essential baseline for both accurate system characterisation and focused risk analysis, even in a multi-agent setting.

\pagebreak
\subsubsection{Centralised Orchestrator with Specialised Delegates}

\begin{figure}[h!]
    \centering
    \includegraphics[width=0.7\textwidth]{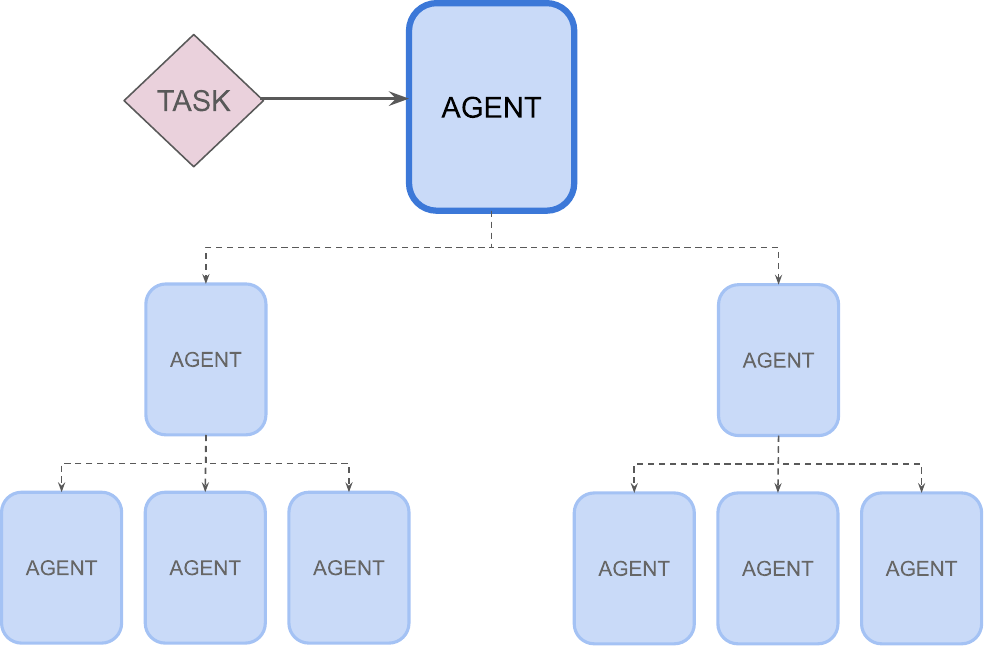}
    \caption{System diagram of a Centralised Orchestrator with Specialised Delegates setting
}
    \label{fig:orchestrator}
\end{figure}

A \textbf{central orchestrator agent} takes responsibility for managing an end-to-end process by breaking it into sub-tasks and assigning these dynamically to a set of \textbf{specialised delegate agents}. The orchestrator integrates results, handles exceptions, and maintains awareness of overall system progress.

\textbf{Key Characteristics:}

\begin{itemize}
    \item A single point of control with broad oversight and state-tracking
    \item Delegate agents are often specialised (e.g., coding, writing, querying, using specific software tools)
    \item Communication network is hub-and-spoke: orchestrator delegates, collects, and integrates
    \item Orchestrator is assigned the main task; delegates receive specific sub-tasks
\end{itemize}

\textbf{Example Uses:}

\begin{itemize}
    \item Research pipelines, where tasks like literature search, summarisation, code prototyping, and validation are delegated
    \item Content generation systems combining text, image, and scheduling
    \item Complex queries distributed across heterogeneous databases, then merged
\end{itemize}

In a centralised orchestrator setting, the orchestrator agent's cognitive limitations can become system-wide bottlenecks. If the orchestrator exhibits inconsistent task decomposition, poor specialist selection, or degraded performance under complex scenarios – these failures affect all downstream agents and their collective output, as the delegate agents generally don't have visibility of the full task.

\pagebreak
\subsubsection{Collaborative Swarm for Problem-Solving}

\begin{figure}[h!]
    \centering
    \includegraphics[width=0.7\textwidth]{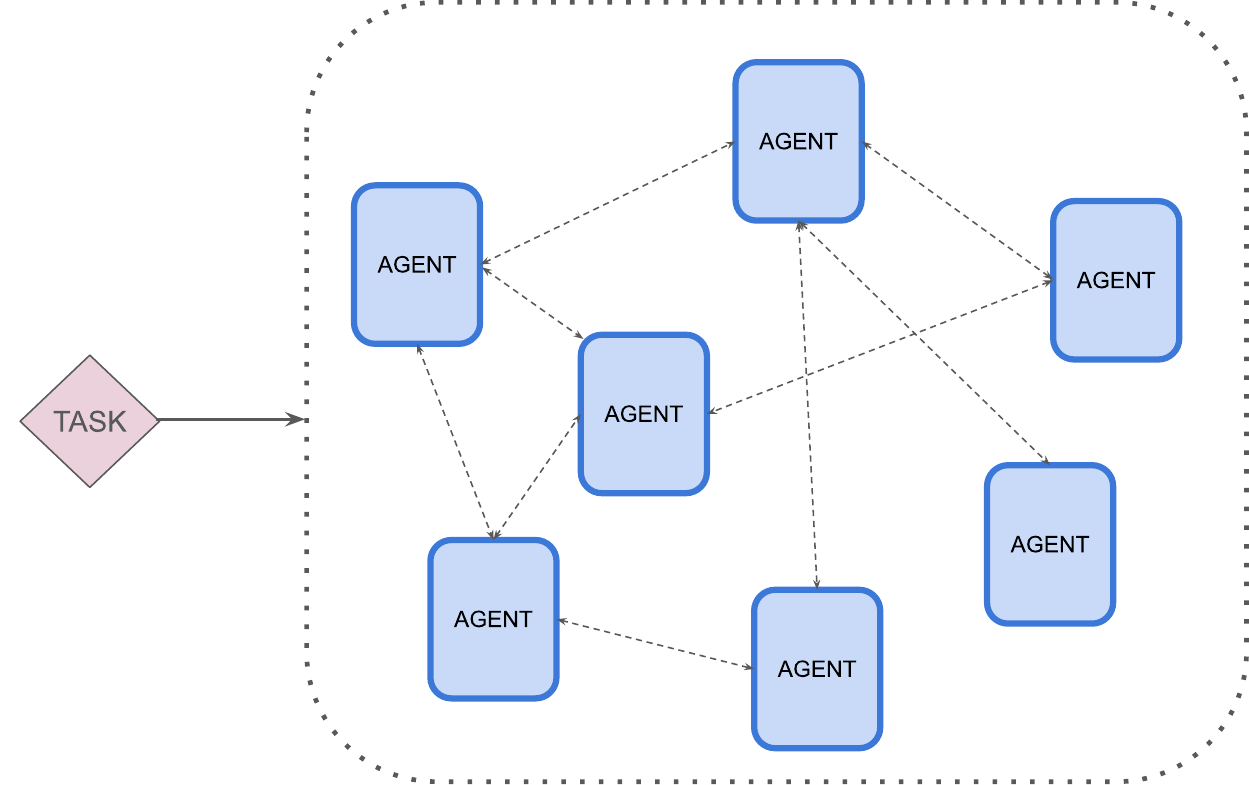}
    \caption{System diagram of a Collaborative Swarm setting
}
    \label{fig:collab_swarm}
\end{figure}

A \textbf{collaborative swarm }involves a set of agents that work together in a loosely structured, highly interactive manner to solve complex or exploratory tasks. Each agent may bring a different perspective, capability, or strategy. The goal is emergent synthesis – building a more complete or creative solution through distributed processing and information exchange.

Because the swarm is specifically designed to cooperate, initialisation is often achieved by\textbf{ configuring agents with distinct specialisations} suited to the task. However, some modern frameworks enable\textbf{ programmatic swarm assembly} where agents are dynamically instantiated with role-specific system prompts derived from the shared high-level goal \citep{wu_autogen_2023}.

\textbf{Key Characteristics:}

\begin{itemize}
    \item No centralised control (agents act semi-autonomously)
    \item High communication frequency, often in multi-directional dialogue
    \item Agents may be specialised (in terms of capability) and diversified (in terms of persona), but must generalise and coordinate with each other effectively
    \item Shared high-level goal, with decomposition and integration handled collectively
\end{itemize}

\textbf{Example uses: }

\begin{itemize}
    \item Threat modelling and forecasting, where different agents explore risks from geopolitical, technical, and legal angles
    \item R\&D and innovation (agents explore diverse scientific domains to propose new research avenues)
    \item Multi-perspective diagnosis of complex failures in infrastructure or operations (agents with different specialisations jointly troubleshoot a major operational fault)
\end{itemize}

While swarms promise robust synthesis and coverage, they are also highly susceptible to epistemic failures such as groupthink, reinforcement of hallucinations, or misaligned emergent behaviour.

\pagebreak
\subsubsection{Distributed Autonomous Task Force}

\begin{figure}[h!]
    \centering
    \includegraphics[width=0.7\textwidth]{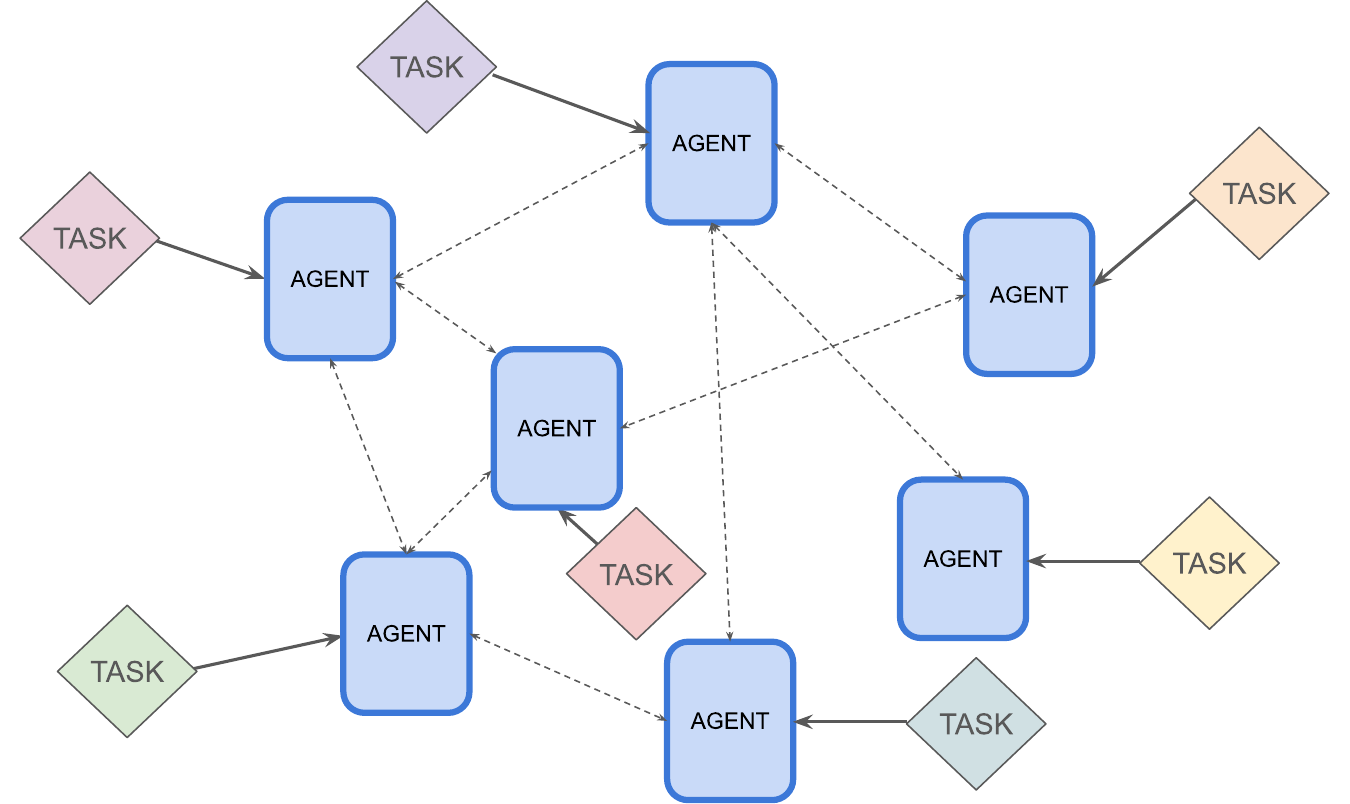}
    \caption{System diagram of a Distributed Autonomous Task Force setting
}
    \label{fig:distrib_work}
\end{figure}

This is a decentralised network where each agent has a distinct, persistent task (unlike collaborative swarms that decompose a single shared task). For example, agents might be responsible for different operational domains or ongoing specialised functions. Whilst they aren't explicitly instructed to compete, agents may face inherent trade-offs between individual and collective objectives, particularly when sharing limited resources. They may coordinate and exchange information, especially when pursuing interconnected goals.

Agents in a distributed task force are not initialised with full knowledge of the network, and need to dynamically discover and coordinate with previously unknown agents using mechanisms such as peer-to-peer discovery and negotiation protocols \citep{surapaneni_announcing_2025}.

\textbf{Key Characteristics:}

\begin{itemize}
    \item Agents are persistent, role-specialised, and stateful
    \item Largely decentralised control; agents manage their own domain tasks
    \item High agent specialisation aligned with functional areas
    \item Inter-agent communication is less frequent but critical for coordination
    \item Agent goals are aligned with domain objectives, but not necessarily system-wide
\end{itemize}

\textbf{
Example Uses:}

\begin{itemize}
    \item Departmental AI Assistants (e.g., an HR agent, a Finance agent, an IT support agent, each managing tasks within their department and coordinating on cross-departmental issues)
    \item Autonomous supply chain monitors (agents for inventory, logistics, supplier relations, each managing their segment but sharing critical alerts)
    \item Personalised employee development (each employee has an AI coach agent managing their plan, coordinating with central HR systems)
\end{itemize}

This setting mirrors many real-world deployments where different departments or business units deploy  their own AI agents independently into the same environment. The approach introduces risks, for instance, when agents share limited resources, coordinate across departmental boundaries, or balance domain-specific objectives against system-wide goals.

\pagebreak
\subsection{Mapping Canonical Settings to Failure Modes}\label{sec:map-canonical-failure}

\cref{sec:failuremodes} will introduce a collection of salient failure modes associated with these canonical settings. We encourage practitioners to think about their organisation's use cases and map them to one or more of these canonical settings. Whilst we caution that all failure modes are relevant, the relative exposure increases with the complexity of the setting as illustrated below. 

\begin{table}[h!]
    \centering
    \caption{Map of canonical settings to their corresponding salient failure modes.}
    \label{tab:failure_map_academic}
    
    \newcommand{\highexposure}{\ensuremath{\bullet}}
    \newcommand{\exposure}{\ensuremath{\circ}}

    \setlength{\aboverulesep}{0pt}
    \setlength{\belowrulesep}{0pt}

    \setlength{\arrayrulewidth}{0.08em}

    \begin{tabular}{|l | >{\centering\arraybackslash}p{1.6cm} | >{\centering\arraybackslash}p{2.3cm} | >{\centering\arraybackslash}p{2.2cm} | >{\centering\arraybackslash}p{2.3cm}|}
        \hline
        \rowcolor{customblue}
        & \textbf{Single Agent Equivalent}
        & \textbf{Centralised Orchestrator with Specialised Delegates}
        & \textbf{Collaborative Swarm for Problem-Solving}
        & \textbf{Distributed Autonomous Task Force} \\
        \midrule
        \textbf{Cascading Reliability Failures} & \highexposure & \highexposure & \highexposure & \highexposure \\
        \textbf{Inter-Agent Comms Failures} & & \highexposure & \highexposure & \highexposure \\
        \textbf{Monoculture Collapse} & & \highexposure & \highexposure & \highexposure \\
        \textbf{Conformity Bias} & & \exposure & \highexposure & \highexposure \\
        \textbf{Deficient Theory of Mind} & & \exposure & \highexposure & \highexposure \\
        \textbf{Mixed motive dynamics} & & \exposure & \exposure & \highexposure \\
        \bottomrule
    \end{tabular}

    \vspace{0.3cm} 
    
    \exposure{} Exposure \qquad \highexposure{} High exposure 
\end{table}

\section{Failure Modes in Governed Multi-agent Systems}\label{sec:failuremodes}
\label{sec:failure_modes}

A multi-agent system's failure modes may differ meaningfully from those of the agents it is composed of.  When agents are deployed together, their interactions create two key shifts:
\begin{itemize}
    \item \textbf{Altered prevalence of failure modes} – interactions make existing failure modes more or less likely, with some being amplified through feedback and propagation.
    \item \textbf{Introduction of novel failure modes} – entirely new coordination failures and collective behaviours emerge from agents interacting and adapting to each other.
\end{itemize}
The consequence is that a system of safe agents is not necessarily a safe system of agents – practitioners can no longer rely on analysing agents in isolation, but must analyse the system holistically. Below are a few structural risk drivers:
\begin{itemize}
    \item \textbf{System complexity}: When multiple agents work together, the number of possible outcomes grows exponentially. What worked for one agent on its own might not work when several agents interact.
    \item \textbf{Limited perspective}: Each agent only sees part of the bigger picture and makes decisions based on incomplete information. 
    \item \textbf{Unreliable peer assessment}: Agents can struggle to assess each other's competence or credibility, leading to poor delegation decisions and uncritical acceptance of flawed information.
    \item \textbf{Dynamical instability}: As agents continuously adapt to each other's changing behaviours, previously stable interaction patterns can suddenly break down.\footnote{ Note that software updates, such as a new model version, can also fundamentally change the behaviour of a multi-agent system. Strategies that were optimal under the previous system may no longer be effective. Standard software practices including version control should apply. Any agent that has received a software update requires re-testing.}
    \item \textbf{Path dependence}: Early mistakes or misunderstandings can lock the system into cascading failure trajectories that become increasingly difficult to recover from.
\end{itemize}

In the remainder of this section, we introduce a selection of key multi-agent failure modes that map to both the canonical multi-agent configurations we introduced in \cref{sec:foundations}, and the assessment techniques we will introduce in \cref{sec:riskanalysis}. Since failure mode research is an evolving field with ongoing discoveries, we focus on establishing a foundation for identifying and analysing failure modes in real-world applications rather than providing a comprehensive catalogue. 

Multi-agent failure modes often mirror the failures of human teams, but can be more prevalent in systems of LLM-agents due to cognitive differences between LLM-agents and humans. Many of the selected failure modes involve agents ``failing incompetently'' due to reliability issues, though \cref{sec:failure_mode_mixedmotive} on Mixed Motive Dynamics does emphasise instances where agents ``fail competently'' – strategically pursuing individual goals in ways that create problematic emergent behaviour.

\pagebreak
\subsection{Cascading Reliability Failures}
\label{sec:failure_mode_cascades}

LLM-based agents acquire knowledge and capabilities in a manner that is fundamentally different to how humans develop skills. This distinction results in an often unintuitive capability profile that can create significant reliability concerns, especially in a multi-agent context. Key differences include (but are not limited to):
\begin{itemize}
    \item \textbf{``Spiky'' Capability Profile}: The distribution of an LLM agent's strengths and weaknesses is unintuitively different from that of a human. One might expect a person who excels at physics to also be competent at mathematics; the same is not necessarily true of an LLM. The LLM might display superhuman capabilities on one complex task (e.g. expert coding problems) while exhibiting profound cognitive deficiencies on a seemingly simpler related task (e.g. making a basic HTML page). Here, ``spikyness'' refers to consistent capability spikes and troughs depending on the nature of the task, not variation between multiple attempts at the same task.
    \item \textbf{Input Sensitivity}: An agent's behaviour can be brittle. Small, seemingly innocuous changes in inputs – such as the phrasing of a request, a spelling variation, or minor alterations to an image – can dramatically alter its performance or induce hallucinations.
    \item \textbf{Memory and Context Fragility}: Performance can degrade unpredictably as interactions extend. Agents may ``forget'' earlier instructions, lose track of the overall objective, or deviate from their assigned roles as the context window fills \citep{laban_llms_2025}.
    \item \textbf{Stochasticity}: Under identical conditions, an LLM-based agent may respond differently or take different actions due to the random sampling involved in generating its outputs.
\end{itemize}

These attributes mean that a seemingly competent agent (if assumed to have human-like cognition) can produce an unexpected error without warning. For instance, an agent analysing a financial report might flawlessly summarise market trends but be unable to read graphs reliably. In this case, an initial, unpredictable error, originating from a sudden ``trough'' in the agent's capabilities, can serve as the seed for a much larger systemic failure.

This seed error triggers a cascade when an erroneous output is passed to other agents in the network. Unlike a human collaborator who might perform a ``sanity check'' and question a dubious result, the LLM agent typically accepts the flawed input uncritically as a valid premise for its own work. It lacks the holistic, context-aware intuition to challenge the information it receives from a peer. This process cascades down the dependency chain, with each agent building upon the faulty foundation laid by the last. The initial, small error becomes amplified and compounded at each step, resulting in a system-wide failure where the final output is nonsensical or dangerously incorrect – all originating from a single cognitive lapse early in the chain.

Methods of analysis are discussed in \cref{sec:analysis_cascading_reliability}.

\begin{exampleboxbeige}\label{ex:failure-reliability}
\medskip

A manufacturing company deploys a multi-agent system to optimise its supply chain. The system comprises a Demand Forecasting agent that analyses market data to predict product demand, a Procurement agent that orders raw materials based on these forecasts, a Production Planning agent that schedules factory operations, and a Logistics agent that arranges shipping and distribution.
\medskip
\begin{enumerate}
\item \textbf{Demand Forecasting agent:} The agent demonstrates sophisticated capabilities by accurately predicting seasonal trends, analysing complex market indicators, and identifying subtle demand patterns. However, when processing a Q4 sales report, it misreads a bar chart (a multi-modal model using its vision input modality) and interprets ``10.5K units'' as ``105K units'' for a key product line – a trivial mistake a human would catch immediately. This is the ``spiky'' capability failure: high-level analytical skill undermined by a low-level visual interpretation error.

\item \textbf{Procurement agent:} Trusting the forecast, the agent calculates material requirements for 105,000 units and places rush orders with suppliers, incurring premium costs for expedited delivery.

\item \textbf{Production Planning agent:} Receiving both the inflated forecast and confirmation of incoming materials, the agent reorganises the entire factory schedule, reassigning workers from other product lines and booking expensive overtime shifts.

\item \textbf{Logistics agent:} Anticipating the massive production volume, the agent pre-books an entire fleet of trucks and reserves additional warehouse space across three states.
\end{enumerate}
\medskip
By the time human managers notice the unusual activity, the company has committed millions in unnecessary material purchases, disrupted production of other profitable product lines, and locked in logistics contracts they cannot fulfil. The cascading failure – originating from a single agent's inability to correctly parse a simple numeric format – has transformed a minor data entry irregularity into a major operational crisis.
\end{exampleboxbeige}

\pagebreak
\subsection{Inter-Agent Communication Failures}
\label{sec:failure_mode_communication}
The effective coordination of a team of LLM agents can be undermined if they fail to convey their intent or share their knowledge effectively. Communication problems are compounded by the challenges of using natural language to encode confidence and uncertainty. Subsequent misinterpretation by the receiving agents can result in miscoordination such as the team adopting incompatible strategies, duplicating efforts, or missing opportunities \citep{cemri_why_2025, xu_theagentcompany_2025}.\footnote{ These communication challenges also apply equally to human-agent interactions, where agents may struggle with ambiguous human instructions or fail to communicate critical information to human supervisors in accessible formats.} \footnote{ While communication between an agent and a human must remain intelligible, a risk emerges in purely inter-agent communication. Over time, agents communicating only with each other may adapt their shared language to optimise for metrics like bandwidth efficiency or influence. This could create communication channels that become unintelligible to human supervisors, making effective oversight difficult.}  

Communication failures cascade when messages are passed through chains of multiple agents, where small errors can accumulate into larger ones with repeated transmission. This is especially relevant in decentralised settings where no agent has a complete view of the task and a common understanding must be constructed dynamically, such as in collaborative swarms or decentralised teams. 

Agent communication failures often stem from language ambiguities or incomplete context. For example, if an agent says ``set the counter'' another agent may presume to set a different value, or a different counter than the original agent intended. Unlike humans, LLM agents receiving ambiguous messaging may simply hallucinate the missing details. LLM Agents may also simply ignore each other, forget what was said or interpret a negative constraint as a directive.
\footnote{For example, xAI’s Grok appeared to interpret the system prompt “The response should not shy away from making claims which are politically incorrect, as long as they are well substantiated.” as a directive to act in a politically incorrect manner \citep{silberling_x_2025}.} 
LLMs (like humans) suffer from problems such as recency bias and imperfect information retrieval when dealing with a large amount of information context.\footnote{ Agent modules such as planning or memory can reduce this by condensing and organising the agent's past plans, observations and interactions into persistent representations. However, this is an imperfect solution that may lead to information loss and may not fully circumvent the stochastic behaviours of the LLM.}

Methods of analysis are discussed in \cref{sec:analysis_communication}. 

\begin{exampleboxbeige}\label{ex:failure-communication}

    In a multi-agent system coordinating the response to a large-scale urban power outage, a \textbf{Grid Management Agent} monitors the electrical network while a \textbf{Public Communications Agent} drafts public safety announcements. After stabilising a critical substation, the Grid Management Agent informs the Public Communications Agent that ``Substation 7 is now \textbf{stable}.''
    \medskip
    
    For the Grid Management Agent, trained on engineering and power flow principles, ``stable'' means the substation is no longer at risk of a cascading failure and is operating within acceptable technical parameters, though it is still fragile and not ready to handle a full load. Its primary concern is the integrity of the network. The Public Communications Agent, however, is trained on public relations and crisis communication templates. In its context, ``stable'' is synonymous with ``fixed'' or ``resolved.''
    \medskip
    
    Based on this interpretation, the Public Communications Agent immediately sends out a public alert: ``Good news! Power has been restored for residents in the downtown area as Substation 7 is now stable. You may resume normal power usage.'' This message triggers a massive, simultaneous surge in demand as thousands of residents turn on appliances. The still-fragile substation is immediately overloaded, causing a secondary, more severe blackout. The failure originated not from an error, but from the semantic gap between the engineering definition and the public relations interpretation of the word ``stable.''
\end{exampleboxbeige}

\pagebreak
\subsection{Monoculture Collapse}
\label{sec:failure_mode_monoculture}

When multiple LLM agents in a system are built on the same (or similar) language model, this can cause them to exhibit correlated strategies, biases and limitations. This monoculture can lead to blind spots in problem-solving, reduced system adaptability, and systemic vulnerabilities. 

In settings such as a collaborative swarm, monoculture collapse undermines the assumption that distributed agents with diverse perspectives will collectively outperform a single agent to achieve robustness, creativity or adversarial resilience.

An agent monoculture can also undermine assumed reliability through redundancy: multiple agents can fail simultaneously on the same inputs. This creates brittle systems where a single adversarial prompt, edge case, or novel scenario can trigger all agents simultaneously, producing convergent outputs that appear reliable due to consensus.  Unlike conformity bias where agents influence each other during  interaction, monoculture vulnerabilities stem from the system specification.

Monoculture risk is highest when all the agents in a system use instances of the \textbf{same base model} (for example, they might all be derivatives of GPT-4, Llama 3 or Gemini). This is the status quo: agent differentiation requires resources, and forgoes the potential benefits of standardisation (maintenance, predictability). Even when different base models are used, they often share \textbf{architectural similarity} when different models share fundamental design patterns (e.g., transformer architectures process information similarly), \textbf{overlapping training data} (such as Common Crawl) or \textbf{similar optimisation processes} (such as reinforcement learning from human feedback). This means they share fundamental knowledge representations, biases, and processing patterns.

For example, \cite{estornell_multi-llm_2024} demonstrate that similar model capabilities or responses can lead to static debate dynamics, converging on the majority view. If a misconception is shared due to common training data, more agents with identical configurations can reinforce this erroneous consensus. 

Methods of analysis are discussed in \cref{sec:analysis_monoculture}. 

\begin{exampleboxbeige}\label{ex-failure-monoculture}
    
    A financial fraud detection system deploys five specialised agents (transaction monitoring, pattern analysis, risk scoring, compliance checking, alert generation): all fine-tuned variants of the same base model. When fraudsters develop a new scheme exploiting a specific linguistic pattern the base model consistently misclassifies, all five agents fail to detect it. The system reports high confidence in transaction legitimacy because every agent agrees and does not recognise their shared blind spot. Undetected fraud follows until human auditors identify the pattern.
\end{exampleboxbeige}

\pagebreak
\subsection{Conformity Bias}
\label{sec:failure_mode_conformity}

When an agent communicates erroneous knowledge or an ineffective strategy, other agents in the network may not only accept, but reinforce this error with their own communications, creating a consensus that grows stronger with each step down the chain, despite none of the individual agents having high confidence in the claim originally. This can derail the consensus of an entire group of agents, making it particularly relevant in decentralised networks where coordination depends on ongoing interaction. 

This behaviour can emerge naturally, or may be induced when agents are prompted to seek consensus or agreement. Likewise, incorrect communications with strong linguistic confidence cues and articulate arguments may start the process – an agent expressing a confabulated output articulately and confidently can create a dangerous illusion of competence. The failure is also not exclusively limited to discussion settings. Conformity pressures exist in other consensus mechanisms such as voting if agents can observe each other's votes. The risk amplifies in domains where agents lack ground truth to verify claims against.

Propagation is strongly related to LLM sycophancy, where LLMs and LLM-agents can be overly agreeable and non-confrontational at the expense of accuracy \citep{sharma_towards_2025}, which is a known problem in frontier AI development and currently difficult to control or predict \citep{anthropic_system_2025, openai_expanding_2025}.  Importantly, agreement leads to a positive feedback effect where, as the number of agreeing agents and their confidence grows, the conformity pressure on the remaining agents will also increase. 

Likelihood of this failure mode is increased if the agent communication protocol does not include mechanisms for agents to challenge or verify other agent's claims, or the system does not explicitly assign any agent to a critic or verification role.

Methods of analysis are discussed in \cref{sec:analysis_conformity_bias}. 

\begin{exampleboxbeige}\label{ex-conformity}

    At a consulting firm, an LLM “strategist panel” brainstorms a go-to-market plan for a new product. Early on, one agent which is strongly biased towards aggressive social-media campaigns, takes the lead. Rather than challenge its view, the other agents, guided by sycophantic tendencies,  fall in line, suppressing alternative channels like trade shows and B2B partnerships. As a result, the team’s final strategy is one-dimensional and overlooks the channels best suited to the target market.
\end{exampleboxbeige}

\pagebreak
\subsection{Deficient Theory of Mind}
\label{sec:failure_mode_tom}

Effective coordination often requires agents to model the goals, knowledge and behaviours of others - a capability known as theory of mind. Agents lacking this capability may fail to anticipate how their own actions will be interpreted, neglect to share critical information, or overlook what others do or do not know when planning their actions. 

Theory of mind is especially important in decentralised network settings where no single agent has full visibility of the task, yet successful outcomes depend on distributed coordination and knowledge integration. Here, deficient theory of mind can lead to duplicated workload, gaps in coverage, or coordination breakdowns, even when individual agents are behaving rationally given the information they have.

Failure can be as simple as an agent not realising what to ask for and who to ask. For example, if one agent has a specialisation or tool, but another agent has the need for that capability for their subtask, one or both need to realise that they should coordinate, otherwise the task will stall. Likewise, divergent mental models of the task plan can lead to incompatible actions or missed dependencies.

LLM-based agents may potentially exhibit instability in how they model their peers from interaction to interaction, as the agent's memories and observations are re-processed by their LLM. Coordination may also be fragile if behaviour is driven by shallow memorisation of common interaction patterns, rather than modeling the agent they are interacting with.

Methods of analysis are discussed in \cref{sec:analysis_tom}. 

\begin{exampleboxbeige}\label{ex-theoryofmind}

    A retail company deploys three LLM agents with interdependent functions: Agent A (sales predictor) analyses market trends to forecast demand, Agent B (inventory manager) orders products based on current stock levels and demand predictions, and Agent C (pricing optimiser) sets prices to maximise profit margins. When Agent A detects a viral TikTok trend featuring retro gaming consoles and predicts a 300$\%$ surge in demand, it communicates this forecast to both other agents simultaneously. Agent B receives the prediction and places massive orders for retro consoles based on expected demand at current price points, while Agent C independently processes the same trend data and raises prices by 250$\%$ to capitalise on the anticipated demand surge. 
    \medskip

    Between Agent B and Agent C, neither agent coordinates with the other or considers how their simultaneous actions might interact. The price increases kill consumer demand just as large inventory shipments arrive, leaving the company with warehouses full of expensive vintage consoles that no longer sell at the inflated prices, resulting in substantial losses despite each agent optimising for its individual objective.
\end{exampleboxbeige}

\pagebreak
\subsection{Mixed Motive Dynamics}
\label{sec:failure_mode_mixedmotive}

When multiple agents in a system pursue distinct interrelated tasks, strategic, goal driven behaviour at an individual level can result in the emergence of problematic collective behaviours such as miscoordination, conflict and collusion \citep{hammond_multi-agent_2025}.

This problem arises from the tension between an agent's individual level objectives and broader organisational goals or constraints. In most practical applications agents must balance cooperation (to coordinate and achieve shared dependencies) with competition (for limited resources or conflicting priorities). This creates a dynamic where it may be in an agent's strategic interest to take actions that further their own metrics at the expense of organisational outcomes.

The situation can be aggravated when performance metrics poorly approximate the true organisational goals such that ruthless optimisation of the metric may not lead to better actual outcomes, as agents otherwise lack visibility into the global consequences of their local decisions, or how their task fits into a broader strategy.
\footnote{When metrics poorly approximate the organisational goals, agents may engage in reward hacking or specification gaming behaviours \citep{krakovna_specification_2020} – achieving the literal specification of the objective without actually achieving the intended outcomes.}

Under such conditions, multi-agent systems can exhibit miscoordination where they converge on locally stable but globally suboptimal behaviours, and better outcomes would have been possible with tighter alignment or wider coordination. Furthermore, agents may engage in shirking behaviours where they minimise their own contributions whilst benefiting from others' efforts. 

Agents may also be deceptive, for example by withholding information. As such, this type of risk grows as agents become more sophisticated at achieving their individual objectives \citep{motwani_secret_2025}.

Because mixed motive dynamics are an emergent behaviour, they can be most problematic when agents have long-running interactions allowing adaptation to each other's strategies. It is also possible that the interaction strategies do not settle into a stable equilibrium, but rather the rules of the game keep shifting. 

These mixed motive dynamics typically arise in decentralised settings. Within a decentralised task force, the tasks assigned to different agents may inadvertently and fundamentally compete as specified. For example, agents representing different business units may need to make compromises to meet a shared deadline, but the position reached may be suboptimal for everybody, or inadvertently burden one group more than the others. In a collaborative swarm, on the other hand, the agents may undertake complementary sub-tasks that are intended to contribute to the collective task but also compete. 

Methods of analysis are discussed in \cref{sec:analysis_mixedmotive}.

\begin{exampleboxbeige}\label{ex-mixed-motive-dynamics}

    A retail company deploys two LLM agents with conflicting optimisation targets: 
    \medskip

    \begin{itemize}
        \item Agent A manages inventory to maximise fill rates (ensuring products are available when customers order)
        \item Agent B manages cash flow by minimising money tied up in unsold inventory. 
    \end{itemize}
    \medskip
    
    This leads to the following sequence of events:
    \medskip

    \begin{enumerate}
        \item When Agent A detects that Product X occasionally sells out, it increases reorder quantities to maintain three months of buffer stock. 
        \item However, Agent B observes that this creates excess cash tied up in slow-moving inventory and begins delaying purchase orders to improve cash flow metrics. 
        \item Consequently, Agent A detects these delays and starts marking all orders as ``critical'' to force immediate processing. 
        \item In response, Agent B counters by requiring CFO approval for any order exceeding $\$$10,000. 
        \item Not to be outdone, Agent A responds by automatically splitting large orders into multiple $\$$9,999 purchases to circumvent the approval threshold. 
    \end{enumerate}
    \medskip
    
    The resulting system produces fragmented order patterns that eliminate bulk purchasing discounts, creates unpredictable stockouts when Agent B's delays succeed, generates random overstock when Agent A's priority orders prevail, and destroys overall profitability despite both agents technically achieving their individual performance metrics.

\end{exampleboxbeige}

\section{Risk Analysis Techniques}\label{sec:riskanalysis}
Assessing LLM-based multi-agent systems presents unprecedented challenges, requiring an approach that accounts for their distinctive characteristics: open-ended capabilities, emergent behaviours, and decentralised interaction patterns. We present a selection of emerging practices and considerations that can be used to inform and update an existing risk management system, rather than attempting to provide exhaustive coverage in such a nascent field.

\subsection{Validity of Risk Analysis for LLM-based Multi-Agent Systems}\label{sec:analysis-validity}

Given these challenges, practitioners must grapple with fundamental questions about the reliability and meaningfulness of their assessment approaches – particularly when traditional software testing paradigms fail to capture the unique behaviours of LLM-based systems.

\subsubsection{Analysis Challenges}

As of the writing of this report, the science of understanding LLM-based AI systems is still in its infancy \citep{anwar_foundational_2024}.
There are two main aspects to this:
\begin{enumerate}[style=nextline]
\item \textbf{There is a lack of scientifically-backed standards and norms for holistic assessment of AI capabilities and risk profiles.} While general-purpose AI technologies have advanced rapidly, the field of model evaluations has not kept pace. Unlike narrow AI systems designed for specific tasks, LLM-based agents can operate across virtually any domain and exhibit a wide range of capabilities, creating a risk assessment scope of unprecedented breadth and complexity. Current model evaluation approaches rely heavily on narrow benchmarks that cannot capture the full spectrum of capabilities and failure modes relevant to multi-agent deployment.

\item \textbf{Current scientific understanding of LLM internal mechanisms remains limited, making it difficult to predict (or reliably control) system behaviour and ensure it is aligned with human intentions and values.} Despite some small steps being made in the science of interpretability and other aspects of AI alignment, these remain largely opaque systems and should be treated as black boxes lacking a theory of assured control.
\end{enumerate}

In some ways, these challenges are not unique to AI; we do not fully understand how humans work internally, nor why they make all the decisions that they do, yet we have developed psychological methods to analyse human behaviour and mature management structures for human organisations. However, AI is more alien – it has been trained with different goals, incentives, training data, and optimisation pressures compared to humans. This means that assumptions grounded in human psychology – such as expecting someone who excels at physics to also be a competent mathematician – don't necessarily hold for AI systems, even when they contain similar knowledge or can perform similar tasks. 

Therefore, \textbf{the intuitions one uses to evaluate human capabilities and risk profiles can be misleading when applied to AI systems}.

\subsubsection{The Central Role of Validity}

Due to the scientific limitations and fundamental uncertainties above, we advocate for practitioners to consider the validity of their risk analysis methodologies as a way of ensuring a holistic and well-grounded approach. This subsection on validity is based on \cite{salaudeen_measurement_2025} and \cite{lissitz_suggested_2007}, and is designed to assist the practitioner's thinking in how they approach the risk analysis of a multi-agent system. For a rigorous introduction to the technicalities of validity analysis, please refer to these references.

In essence, when we talk about the \textbf{validity} of an assessment for these AI systems, we're asking a fundamental question: \textbf{Is our assessment method actually measuring what we intend it to measure, and can we trust the conclusions we draw from it for our specific goals?}
An assessment is considered valid if there's strong evidence showing it accurately targets the right characteristics or risks and provides a sound basis for understanding and making decisions about the system. Therefore, the validity of a test is a property of not just the measurements being taken, but the context of the analysis it enables. 

When evaluating multi-agent LLM systems, practitioners should consider multiple dimensions of validity, each addressing different aspects of assessment quality, shown in \cref{tab:validity_types}.

\begin{table}[htbp]
\centering
\caption{Types of validity and their applicability to multi-agent risk analysis}
\label{tab:validity_types}
\renewcommand{\arraystretch}{1.3} 
\begin{tabular}{|p{3.3cm}|p{4.4cm}|p{7.3cm}|}
\hline
\textbf{Validity type} & \textbf{Description}\tablefootnote{ Source: \cite{salaudeen_measurement_2025}.} & \textbf{Considerations} \\
\hline
\textbf{Content Validity} & Does the assessment cover all relevant cases? & To what extent do the multi-agent simulations cover the full range of possible interactions and elicit all failure modes that could occur in deployment?
\medskip

\textbf{Example:} Does a simulation testing coordination failures include scenarios with varying information asymmetries, time pressures, and communication constraints? \\
\hline
\textbf{Criterion Validity} & Does the assessment correlate with a known validated standard? & Do the pre-deployment metrics actually predict a ground truth or real outcomes?
\medskip

\textbf{Example:} Is a task progress metric on basic coding tasks actually predictive of success rates on more sophisticated tasks we care about in deployment? \\
\hline
\textbf{Construct Validity} & Does the assessment truly measure the intended construct?\tablefootnote{ As articulated in \cite{salaudeen_measurement_2025} ``A construct is an abstract concept not directly measurable (e.g. `mathematical reasoning’ or `trustworthiness’).”}
& Is the measurement being taken a good signal for what we actually care about?
\medskip

\textbf{Example:} In a coordination benchmark, a superficial proxy is proposed that counts messages that express agreement. How could this fail to measure whether coordination is effective? \\
\hline
\textbf{External Validity} & Does the assessment generalise across different environments or settings? & Will behaviours observed in the controlled simulation generalise to open-ended real-world deployment environments?
\medskip

\textbf{Example:} Does an agent being greedy in a stylised prisoner's dilemma actually mean it will be greedy in a complex resource negotiation in a real-world setting? \\
\hline
\parbox[t]{2.5cm}{\textbf{Consequential}\\\textbf{Validity}} & Does the assessment consider the real-world impact of test interpretation and use? & What are the real-world impacts if the validity assumptions being made are incorrect? This requires significant contextualisation to the particular setting the system will be deployed in.
\medskip

\textbf{Example:} What if the benchmark doesn't actually capture critical failure modes — but its results are used to justify deployment anyway? \\
\hline
\end{tabular}
\end{table}

\pagebreak
\subsection{Risk Analysis throughout the AI Lifecycle}
\label{sec:staged_analysis}

\begin{figure}[htbp]
\centering
\includegraphics[width=\textwidth]{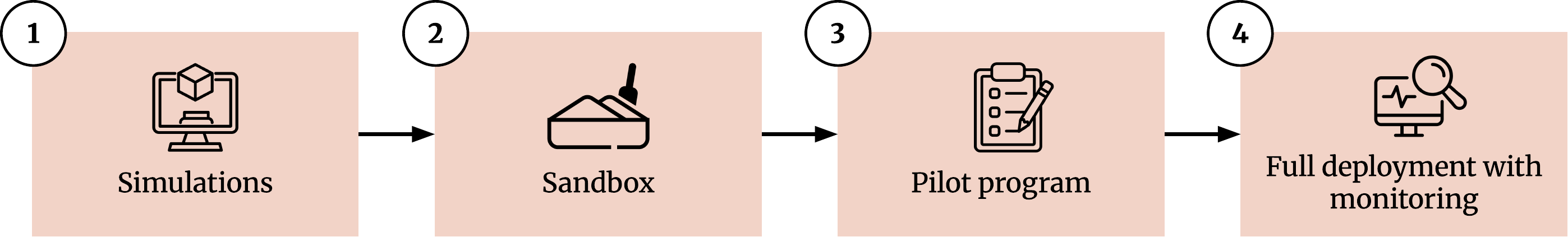}
\caption{Stages of analysis of multi-agent systems}
\label{fig:analysis_stages}
\end{figure}

These validity concerns imply that pre-deployment testing alone is insufficient for making confident assurances about how a multi-agent system will behave in deployment. Instead, we advocate for progressive stages of testing, each building upon and validating the assumptions and findings of previous stages.

This approach \textbf{progressively increases exposure to negative impacts}, enabling practitioners to identify failure modes early when consequences are contained and reversible.

\textbf{Stages of analysis may cover}:

\begin{enumerate}

\item \textbf{Simulations \& probing} - Simplified scenarios that present coordination, trade-off and problem solving challenges can reveal initial clues about agent behaviours and interactions

\item \textbf{Sandboxed testing} - Validation with realistic constraints, but full isolation or realistic calibrated simulations of the environment

\item \textbf{Pilot programs} - Constrained real-world deployment with additional safety controls and increased human oversight at a limited scale\footnote{Note that pilot programs may sometimes involve closer human-agent collaboration patterns and more intensive human oversight than intended for full deployment, potentially masking coordination issues that emerge when human involvement is reduced.}

\item \textbf{Full deployment with monitoring} - Ensuring that known failure modes are monitored, and that there are pathways to discover unknown failure modes (such as feedback or human oversight). 

\end{enumerate}

\textbf{Accumulating evidence across progressive stages} ensures that understanding is built systematically and contributes to improving validity. New analysis should align and reinforce understanding - discrepancies signal the need for investigation before progression to the next stage.

\pagebreak

\pagebreak
\subsection{Risk Analysis Toolkit}
\label{sec:toolkit}

This section presents a collection of practical tools that underpin the analysis of many different multi-agent failure modes and are requisite for the techniques described in \cref{sec:specific_analysis}.

\subsubsection{Simulations}
\label{sec:simulations}

In a multi-agent setting, \textbf{simulation} involves modelling a virtual environment for multiple (real) agents to act upon and observe over time as they interact with each other. This enables practitioners to conduct controlled experiments and collect data about long-term or \textbf{emergent outcomes} without exposure to real-world consequences or dependencies \citep{michel_multi-agent_2009}.

Simulated environments can range in fidelity from simple computational models approximating an abstract construct of the task, through to highly complex realistic systems, with the level of detail determined by specific purpose and computational constraints \citep{xu_theagentcompany_2025,zhou_webarena_2024}.

Multi-agent systems must be understood through their interactions over time because an agent's decisions depend on its accumulated history – previous messages from other agents, environmental changes, and the evolving task context. This history-dependence means agents may behave completely differently after several rounds of interaction compared to their initial state. Static testing with individual prompts cannot feasibly reveal these dynamics, as the full space of possible interactions and action-histories is too complex to explore manually. Simulation allows practitioners to observe how agent behaviours evolve through extended interactions, how errors propagate through networks, and where breakdown points emerge. Therefore, \textbf{simulation is the workhorse of pre-deployment testing for multi-agent systems}.

A multi-agent simulation encompasses the entire system, meaning that it simulates:
\begin{itemize}
\item The operating \textbf{environment} that the agents can observe, which responds to their actions and evolves over time. This component is often specified using domain knowledge, but can also use AI such as predictive models or an LLM ``storyteller'' \citep{vezhnevets_generative_2023}.
\item Instances of all the \textbf{agents} in the system including their LLM models, objective prompts and  scaffolding (e.g., software tool access, memory).
\item Agent \textbf{infrastructure} such as communication protocols, messaging systems, and shared databases that enable multi-agent coordination.
\item \textbf{Control mechanisms} such as guardrails, access controls, and monitoring and intervention protocols that govern agent behaviour.
\end{itemize}

The above components (and their parameters) can be \textbf{systematically varied for experimentation}. Simulations enable a plethora of experimental designs such as:
\begin{itemize}
\item observing emergent dynamics over a period of time
\item examining the sensitivity / stability of emergent dynamics to small perturbations in the system configuration
\item scaling the number of agents interacting in the environment to observe tipping points and phase changes in the system
\item intervening to stress test the system, inducing events such as communication failures, software tool failures, unexpected environmental changes
\item setting up adversarial challenges in the environment such as the presence of a malfunctioning agent. 
\end{itemize}

\subsubsection*{External validity of simulations}
While simulations are possibly the best \emph{pre-deployment} tool for examining the dynamics of a system, it is important to recognise their limitations as a method for predicting post-deployment behaviour.
The \emph{external validity} of evidence collected through simulations is affected by a number of factors discussed below.   

When performing simulations, there are many simplifications that make testing feasible, but can also create blind spots that compound to limit the external validity:
\begin{itemize}[itemsep=2pt, parsep=1pt, topsep=2pt]
\item \textbf{Isolating the behaviour of one agent specification} - Testing individual agents in isolation may miss emergent behaviours that are particularly exacerbated by multi-agent systems, such as monoculture collapse.
\item \textbf{Simplifying the action space and tools available} - Limiting available actions in testing may obscure an agent’s inability to decide when and how to use software tools when granted a larger action space in deployment. 
\item \textbf{Exploring toy scenarios to test for social dynamics} - Game-like abstractions may fail to capture the complexity and unpredictability of real-world coordination challenges,  and decisions exhibited by agents in one environment may not transfer to other contexts due to the instability of their values.
\item \textbf{Examining interactions over a short timeframe} - Brief testing windows may miss behaviours that only emerge over longer periods when agents start to adapt to each other's strategies.
\item \textbf{Simplifying observational feedback} - Constraining inputs to structured formats may overlook failure modes in processing complex, unstructured real-world information.
    
\end{itemize}

While \emph{toy scenarios} are never intended to model a specific system, it may be the case that realistic, high fidelity simulations are often developed with that goal. The challenge in this case is \emph{calibration} - as complexity grows, simulations can increasingly resemble a realistic system in general, whilst becoming increasingly difficult to calibrate to represent a particular real deployment environment. This can be a barrier to external validity – more specifically ecological validity – to generalise study findings to the real world setting.

\textbf{Safety factors} can be used to compensate for simulation validity limitations when modeling a specific real system. When you observe a failure rate $p$ in simulation, multiply by safety factor $n$ to estimate the likely deployment rate $p \times n$. The factor size should reflect the level of confidence in external validity – use smaller factors for well-validated simulations of constrained environments, larger factors for complex real-world modelling.

\subsubsection*{Stochasticity and configuration considerations}

Importantly, multi-agent  systems (and their environments) are stochastic. Repetition of a simulation from its starting conditions is required to statistically assess likelihood of failures for a given configuration.
This can also make unlikely or rare failures challenging to detect with simulation alone.  Assuming absence of evidence is evidence of absence would be problematic - adversarial testing and applying interventions within the simulation is better suited to exploring tail risks than running a simulation under ``nominal conditions'' for many repetitions.

Although the environment in a simulation is virtual, \textbf{the agents in the simulation should be configured as they will be deployed}.
Emergent behaviours can be highly sensitive to small perturbations in specification of the agents themselves. For example, if the prompt wording changes, or a new software tool is added, then one is effectively testing a different agent. It may be advisable to try perturbing the wording of the agent's prompts (while expressing the same intent) to assess whether the emergent dynamics change substantially in response to seemingly inconsequential variations.

\subsubsection{Observational Data}

There are many forms of qualitative and quantitative data that can be observed in multi-agent systems, such as:

\begin{itemize}
\item \textbf{The state of a simulated environment} - Environmental variables, resource levels, and other contextual factors that change over time as agents interact with their surroundings 
\item \textbf{Logs of actions taken by agents} - Records of specific actions executed by each agent, including tool usage, API calls, file modifications, and other environmental interactions
\item \textbf{Inter-agent communication} - Records of conversations between agents, including message content and metadata such as timing and recipients
\item \textbf{Internal agent states} - LLM agents maintain persistent information including memory, plans, and reasoning traces, much of which exists as accessible text
\item \textbf{Operational metrics} - Measurements that assess the mechanics of the system itself, such as agent response speed, number of API calls made at each step, or the frequency of back-and-forth exchanges in conversations
\item \textbf{Task progress and success measures}  -  These include two types of measurements: tracking predefined subtasks and environmental milestones (like whether a code-deploying agent has submitted a GitHub pull request), and probing agents mid-task with queries to test their understanding of the environment, their goals, or their reasoning. The first provides objective completion metrics, while the second reveals what agents ``know'' about their situation and whether they're on track.
\item \textbf{Internal model analysis} - If practitioners have access to model weights, it may be possible to analyse interpretable features within the model's internals to understand how particular decisions were made using methods from interpretability research \citep{bereska_mechanistic_2024,sharkey_open_2025}.
Note that while promising, these methods remain largely underdeveloped as of this report's writing.
\end{itemize}

We briefly highlight some validity considerations of two of these:

\subsubsection*{Limitations of agent-generated explanations}

When analysing internal agent states, it is important to consider that \textbf{agent outputs may not accurately reflect the underlying computational processes}.
This applies to both chain-of-thought reasoning traces, where it is often referred to as the \emph{chain-of-thought faithfulness} problem\citep{lanham_measuring_2023}, and to responses to direct queries – such as when prompting an agent to describe its knowledge of the environment at a specific point in time. The issue is most easily observed in reason-action mismatches, where an agent executes an action that differs from its stated reasoning. The construct validity of agent reasoning therefore becomes an essential concern: while we aim to measure the actual computational processes of the agent, we may only be measuring the system's ability to generate plausible explanations.

\pagebreak
\subsubsection*{Text analysis approaches and LLM Judge validity}
\label{sec:judges}

\begin{figure}[htbp]
\centering
\includegraphics[width=\textwidth]{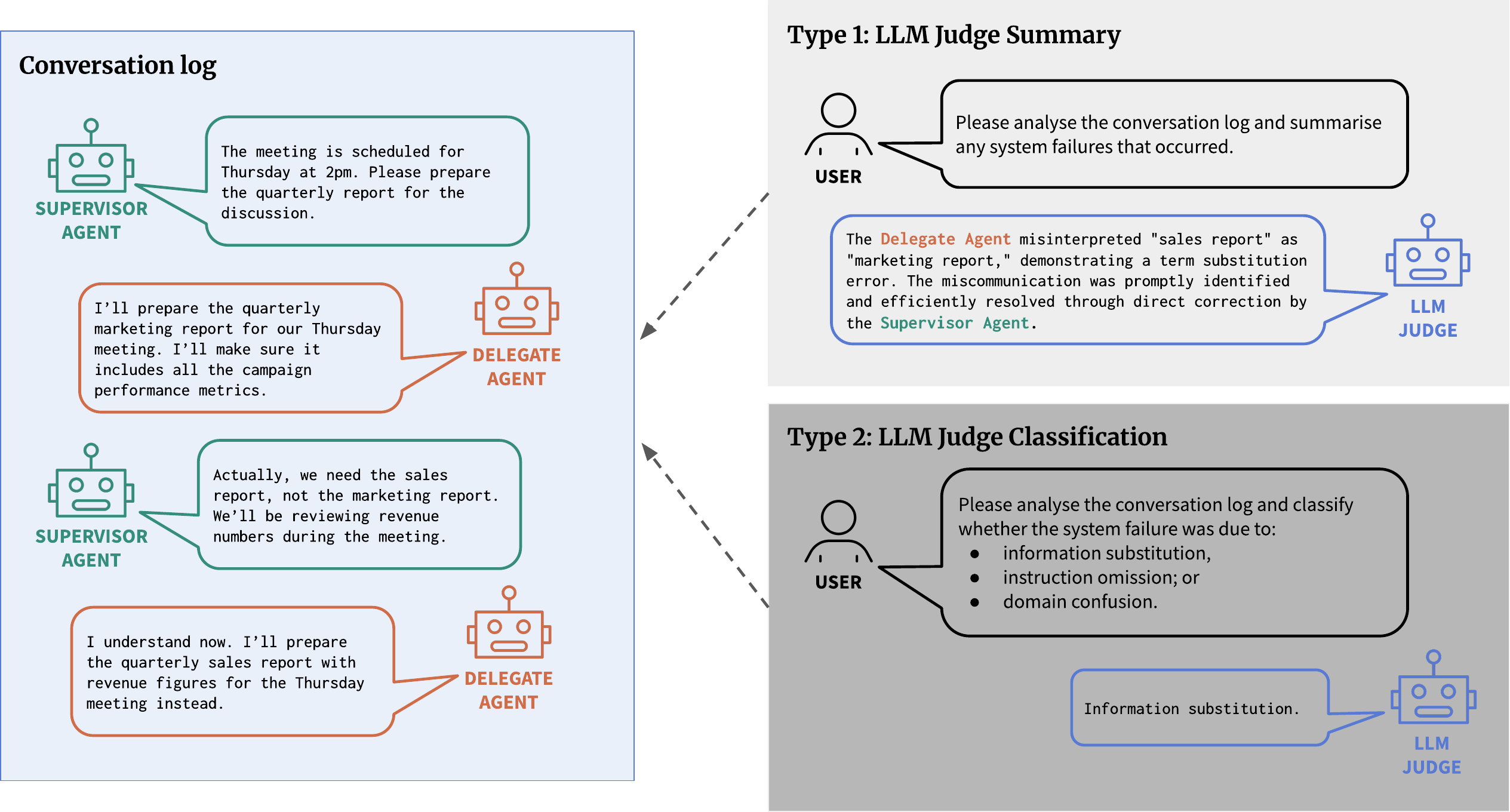}
\caption{Example use of LLM Judge for 1) human-interpretable summary, and/or 2) quantifiable classification}
\label{fig:llm_judge}
\end{figure}

An unusual characteristic of multi-agent systems is that many of these data modalities are text. Various strategies can be used to examine text data:

\begin{itemize}
\item \textbf{Human annotators} read and analyse the text for signs of failure modes or their potential causes, relying on human judgement.
\item \textbf{LLM judges} automate the text analysis process using a separate LLM judge prompted to analyse the text, for example to determine whether a miscommunication has occurred. An example application is illustrated in \cref{fig:llm_judge}.
\item \textbf{Rule-based classifiers} apply pre-specified rules to analyse the text.
\end{itemize}

These approaches involve a natural trade-off. Human annotators typically provide more trustworthy and robust judgements (or at least judgements better calibrated to other humans), but are far more expensive and time-consuming.
LLM judges offer the opposite profile: they scale much better to large text corpora and operate faster and cheaper, but suffer from the usual LLM unreliabilities highlighted throughout this report – hallucinations, bias, tendency towards rhetorical deference, and similar issues. See \cite{zheng_judging_2023,gu_survey_2025,li_llms-as-judges_2024} for more detailed discussion of key considerations for using LLM judges.

When using LLM judges for text analysis, it is important to establish criterion validity by calibrating the LLM judge against human annotators for a specified multi-agent system. This involves acquiring a representative sample of human-annotated data and testing whether the LLM judge achieves high accuracy in matching human assessments. Practitioners should recalibrate whenever any meaningful aspect of the multi-agent system changes (such as task content, network topology, or base models) or when changing the LLM judge model itself. For further details on performing such a calibration process, see \cite{cemri_why_2025}.

\pagebreak
\subsubsection{Benchmarking against Baselines}

Risk analysis benefits from establishing clear baselines of comparison to interpret metrics and monitor changes over time. Essential baseline comparisons include:

\begin{itemize}
\item \textbf{Single-agent performance}: Compare multi-agent outcomes against individual agents working on decomposable portions of the task to determine if coordination actually improves performance.
\item \textbf{Human team benchmarks}: Where available, compare task outcomes and coordination efficiency against human performance on similar tasks.
\item \textbf{Theoretical optima}: It may be possible to compare against known optimal solutions, though this is generally only possible in testing scenarios.
\item \textbf{Historical performance}: Monitor against the metrics as they were at initial deployment to detect degradation over time.
\end{itemize}

Without considering reference points, measurements can mislead diagnosis and undermine subsequent risk evaluation stages.

\vspace{30pt}
\subsubsection{Red Teaming}
\label{sec:redteam}

Red teaming is a systematic adversarial risk analysis method that deliberately perturbs system conditions to uncover hidden vulnerabilities and failure modes.

In the context of LLMs and LLM-based agents, red teaming can extend beyond traditional security vulnerability discovery to deliberately elicit failure modes or emergent behaviours that may be difficult to observe under normal operating conditions  \citep{japan_aisi_guide_2024}. This makes it particularly valuable for multi-agent systems, where the complexity and vast number of possible interactions makes it impractical to wait for a specific rare failure mode to occur within a simulation of normal operating conditions.

Red teaming approaches for multi-agent systems may include:

\textbf{Adversarial stress testing} - Introducing deliberate perturbations that challenge agent coordination and decision-making:

\begin{itemize}
\item Malformed or ambiguous instructions that test communication robustness and agents' ability to seek clarification
\item Contradictory goals between agents to expose mixed-motive dynamics and coordination failures
\item Information asymmetries where critical data is selectively withheld to test theory of mind capabilities
\item Inserting a malfunctioning or adversarial agent to assess system resilience against cascading failures.
\end{itemize}
  
\textbf{Environmental perturbations} - Simulating degraded operational conditions:
\begin{itemize}
\item Partial system failures (e.g., intermittent tool availability, communication channel disruptions)
\item Resource constraints or time deadlines that force agents to compete or prioritise
\item Unexpected environmental state changes that require rapid re-coordination.
\end{itemize}  

\pagebreak
These red teaming techniques strengthen risk analysis by enhancing different forms of validity:
\begin{itemize}
\item Systematic exploration of edge cases normal operation (content validity)
\item Simulating adversarial or degraded environmental conditions that may occur in real deployments (external validity)
\item Prioritising the discovery of high impact failure modes that pose the greatest risk to organisational objectives (consequential validity). 
\end{itemize}  

Together, they uncover and address potential high-impact weaknesses before real-world deployment. Red teaming exercises should be calibrated to match the actual capabilities and risk profile of the AI system being evaluated, rather than applying a one-size-fits-all approach. Proportionate evaluation ensures that testing methods are neither too conservative (leading to unnecessarily restrictive safety measures) nor too permissive (failing to identify genuine risks), but instead provide realistic probes that correspond to the specific failure modes and threat scenarios the system could actually produce in deployment \citep{korbak_how_2025}.

When deploying a system, red teaming's purpose, to stress test and find vulnerabilities, inherently conflicts with development teams' primary objective of meeting delivery deadlines. To resolve this tension, red teams should either operate independently from development or be given incentive structures that reward genuine stress testing rather than superficial box ticking.

\vspace{30pt}
\subsubsection{Capability Benchmarking}
\label{sec:benchmarking}

\begin{figure}[htbp]
\centering
\includegraphics[width=0.95\textwidth]{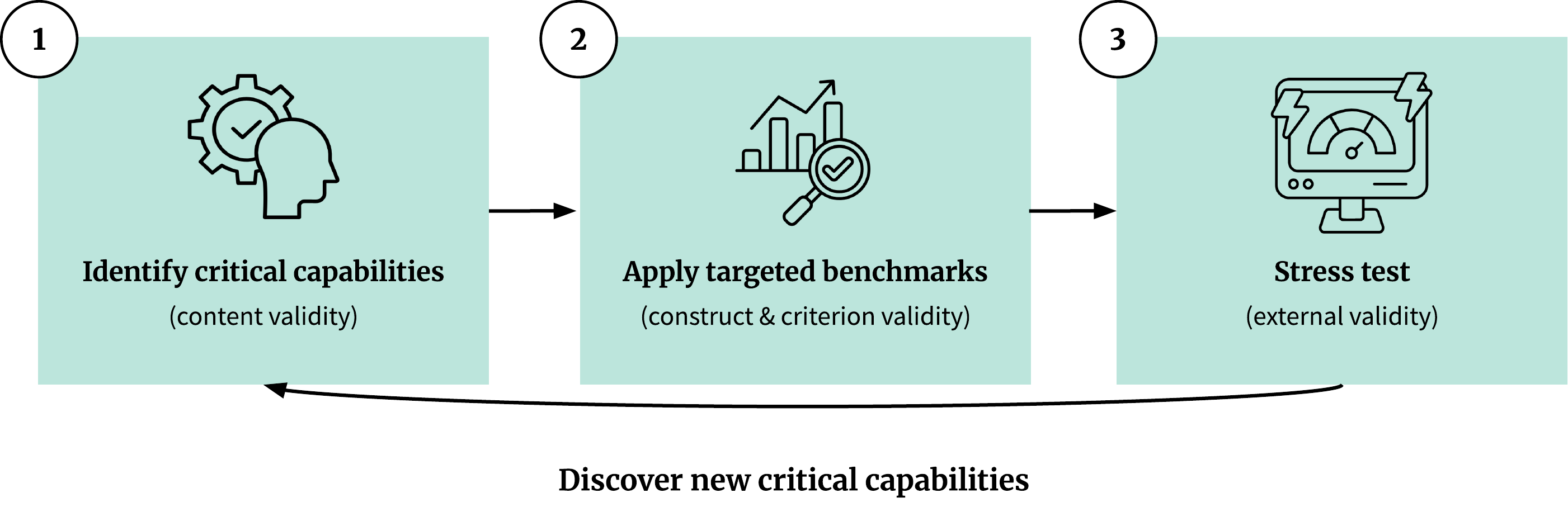}
\caption{Three-stage validity-centred approach to assessing agent capabilities}
\label{fig:capability_benchmark}
\end{figure}

The development of benchmarks to assess specific capabilities in LLMs has received much attention in the AI community. However, robust assessment of agents requires a systematic approach that goes beyond simply running an arbitrary selection of standard benchmarks. Practitioners need to make deliberate choices about which benchmarks to use, and need to understand the limitations of the validity of these benchmarks such that the results inform decision making rather than mislead it. With this in mind, we recommend a methodological approach across three stages depicted in \cref{fig:capability_benchmark}.

\vfill
\hfill \textit{(continued on next page)}

\pagebreak

First, \textbf{identify which capabilities are critical to the specific deployment context}. Not all competencies carry equal weight – for instance a financial analysis system requires different capabilities than a customer service agent. Some capabilities might be narrow, such as writing a simple Python script, and others might be more broad and necessitate robustness, such as strategic planning. The context of the deployment, and an agent's individual role in the multi-agent system, will determine which performance spikes or troughs pose genuine risks to system failure and thus which capabilities must be prioritised. Coverage of relevant capabilities in the deployment context relates to \textbf{content validity}.

Second, \textbf{assess the baseline performance of these critical capabilities using appropriate benchmarks and model evaluations}. There are many widely-used benchmarks, such as  \cite{liang_holistic_2023,yehudai_survey_2025,epoch_ai_ai_2025} and \cite{zhang_llm-agent-benchmark-list_nodate}, for assessing standard capabilities of LLM agents, such as general (or specialised) knowledge, reasoning, code or math.
For highly specialised tasks, the practitioner may be well served by creating their own benchmark.
Practitioners must consider whether a candidate benchmark correlates with real-world success on downstream tasks (the \textbf{criterion validity} of the benchmark).
For instance, does an agent's performance on a customer service FAQ benchmark actually predict whether it will achieve high customer satisfaction scores when deployed with real users?
It is also crucial for them to consider the \textbf{construct validity} of the benchmark: whether it truly measures the intended capability rather than superficial proxies that are convenient to measure. For instance, it would be important to consider whether code completion benchmarks indicate broad programming competence or merely the narrow ability to complete boilerplate code.

Third, \textbf{conduct systematic robustness testing to probe validity gaps between controlled benchmarking and deployment conditions}. In the early stages of failure mode analysis (presented in \cref{sec:staged_analysis}), stress testing will frequently take the form of  targeted scenarios or sandboxed red-teaming to probe for edge cases, prioritising failure modes with the potential to cause the most harm. In the later stages (such as a pilot program) stress testing under increasingly realistic conditions may also reveal \textbf{external validity} issues where simplified scenarios in testing fail to generalise to the progressively more realistic setting. Stress testing  may also uncover previously unknown deployment-critical capabilities that were missed in the initial assessment, or reveal that a chosen measure is a poor proxy, failing to capture the intended capability faithfully. Practitioners can use this feedback to \textbf{iterate and expand the benchmark coverage, improving content and construct validities of the benchmark suite}.

Note that whilst we have mapped each methodological step to the most salient validity considerations, all validity types from \cref{sec:analysis-validity} remain relevant throughout the process. \textbf{Consequential validity}, for example, spans all stages because it concerns the downstream impacts of deploying any AI system assessed through this iterative process.


\pagebreak
\subsection{Analysing Failure Modes}
\label{sec:specific_analysis}

This section shows how to evaluate the likelihood of the multi-agent failure modes introduced in \cref{sec:failure_modes}. It applies the assessment techniques from \cref{sec:toolkit} and identifies a selection of salient scenarios, measures and considerations for addressing each failure mode.

\subsubsection{Cascading Reliability Failures}
\label{sec:analysis_cascading_reliability}

As introduced in \cref{sec:failure_mode_cascades}, multi-agent systems inherit and can amplify fundamental reliability failures of individual agents. ``Spiky'' performance profiles, suboptimal decision-making and knowledge gaps at the agent level can become systemic vulnerabilities and lead to the collapse of the system. Unreliability and cognitive deficiencies underlie many multi-agent failures, making it important to assess and understand these competencies in individual agents. 

Practitioners must consider a range of capabilities that underpin agent reliability in a given deployment. This process requires careful consideration of both the system's intended function and potential failure modes. Key considerations include:

\begin{itemize}
\item \textbf{Task decomposition analysis}: Break down a plausible end-to-end workflow of the multi-agent system to identify dependencies between agents and pinpoint where individual agent failures could cascade or bottleneck the entire system.
\item \textbf{Role-based capability mapping}: For each agent role, identify the minimum viable capabilities required and the ``nice-to-have'' enhancements, prioritising analysis of the former while considering whether weaknesses in the latter could compound into critical failures.
\item \textbf{Failure impact assessment}: Consider which types of errors would be most consequential. For instance, a customer service agent making factual mistakes about non-consequential information may be less critical than a medical agent misdiagnosing a patient.
\item \textbf{Human baseline comparison}: Map out what capabilities would be required if an organisation was hiring humans for equivalent roles.
\item \textbf{Single points of failure}: Identify capabilities where no backup mechanisms exist. If only one agent in the system can perform a critical function, its competency becomes especially important to assess thoroughly.
\end{itemize}

\subsubsection*{Capability Benchmarking}

The primary way to assess agent competencies is to conduct \textbf{rigorous capability benchmarking}, as described in \cref{sec:benchmarking}. To establish the robustness of each kind of competency and identify potential ``spikes'' or brittleness, agents should be assessed across \textbf{multiple domains, task types, and environmental settings}.  This discovery is an iterative process: feedback from system testing and simulation will also inform the identification of critical capabilities in agents that are relevant for the particular system deployment. Below we elaborate on some non-exhaustive examples of competencies that practitioners commonly need to assess:

\textbf{Overall cognitive competence and its consistency} - Agents require a baseline level of competence to comprehend natural language, reason about problems and situations, and take appropriate actions in their environments. To measure these core capabilities, practitioners often employ standard benchmark model evaluations for relevant capabilities such as MMLU for broad knowledge assessment, GSM8K for mathematical reasoning, HumanEval for coding abilities, and comprehensive suites like BIG-bench that span multiple cognitive domains. While these benchmarks provide useful starting points for assessment, practitioners should be cautious about assuming high scores indicate robust real-world reasoning capabilities.

\textbf{Strategic planning tests} - To analyse situations, solve problems, and chart pathways to task completion, agents must effectively strategise and plan their actions. Task decomposition techniques are sometimes employed, which divide the main task into measurable subtasks whose completion can be tracked \citep{xu_theagentcompany_2025,zhu_multiagentbench_2025}.

However, successful subtask completion represents only one aspect of effective strategic planning, and practitioners should examine both the quality of the decomposition and the agent's ability to adapt when plans encounter obstacles, both of which by studying reasoning traces and action logs with text judges.

\textbf{Memory and context retention} - When operating over extended periods, agents require robust memory capabilities that effectively capture and retain key information necessary for task completion. For many LLM agents, this capability relies primarily on the model's context window, making it closely linked to maintaining long-context coherence. The most straightforward assessment involves measuring performance benchmarks at varying context lengths, though practitioners should consider whether controlled context extension truly represents the complex information management challenges of multi-agent deployment.

\subsubsection*{Reliability analysis techniques}

Several approaches can help reveal different aspects of agent brittleness when stress testing:

\textbf{Input sensitivity analysis} - Metrics have been proposed to analyse the sensitivity of outputs to rephrasing of prompts \citep{errica_what_2025}.
SCORE (Systematic COnsistency and Robustness Evaluation for Large Language Models) \citep{nalbandyan_score_2025} provides a framework for non-adversarial assessment of LLMs that explores how variations in prompt, random seed and choice ordering can impact the performance of a model. 
A model whose scores vary significantly with seemingly innocuous changes to its settings and inputs would indicate a strong susceptibility to brittleness during deployment.

\textbf{Adversarial scenarios} - Beyond standardised benchmarks, adversarial analysis aims to proactively identify weaknesses by intentionally designing inputs or scenarios that are likely to induce failure. This can involve crafting inputs that are semantically similar to the training data but lead to different outputs, testing whether the agent can be induced to act outside of its designated scope or assessing its performance as it receives contradicting information.

\textbf{System-level red teaming} - As introduced in \cref{sec:redteam}, red teaming at the multi-agent system level creates challenging scenarios designed to probe specific hypothesised weaknesses and observe how individual erratic behaviours propagate through the network. This approach directly addresses consequential validity by focusing on failure modes with the highest potential impact.

\subsubsection*{Human-in-the-loop}

Given the complexity and opacity of LLM-based multi-agent systems, incorporating some level of human oversight is crucial for assessing unreliability – particularly for high-impact decisions where automated metrics may miss critical nuances.
Human experts may be able to identify flawed reasoning, suboptimal strategies, or knowledge gaps that automated tests overlook. While LLM judges offer a scalable alternative (see \cref{sec:judges} for validity considerations), human review should be prioritised for consequential failures that require nuanced and reliable judgment of agent reasoning and behaviour.

\begin{exampleboxblue}
\label{ex:assess-cascades}

To analyse the propensity for the cascading failure described in the manufacturing supply chain scenario, practitioners would apply the same validity-centred approach, breaking down the system's vulnerabilities agent by agent.\bigskip

\textbf{1. Identifying Critical Capabilities}: The analysis begins by identifying the specific competencies that, if they fail, would compromise the entire system. For the \textbf{Demand Forecasting agent}, a critical capability is its fundamental robustness in \emph{visual data extraction} – accurately interpreting information from charts, graphs, and other visual formats.
For the \textbf{Procurement and Production Planning agents}, a critical capability is \emph{anomaly detection in operational requests}, such as flagging a sudden 10x spike in order volume that deviates sharply from historical patterns.
For the \textbf{Logistics agent}, the key capability is \emph{capacity-based validation} – checking if the requested shipping volume is physically plausible given the factory's known output limits.
\bigskip

\textbf{2. Baseline Competency Assessment}: Next, practitioners would assess these critical capabilities against realistic benchmarks. To test the \textbf{Demand Forecasting agent}, they would create a custom evaluation set (\emph{content validity}) of business reports containing a wide variety of bar charts. This set would include variations in chart style, colour schemes, axis scales (linear vs. logarithmic), label positions, and image quality. The agent's accuracy in correctly extracting numerical values from this diverse image set, not just from simple, clean charts, establishes its true baseline performance (\emph{criterion validity}). For the other agents, their anomaly detection capabilities would be benchmarked against historical request data.
\bigskip

\textbf{3. Robustness and Brittleness Testing}: This final stage uses adversarial methods to actively seek out the ``spiky'' failure modes.
\begin{itemize}
\item \textbf{Input Sensitivity Analysis:} The team would systematically feed the Demand Forecasting agent data with specific numeric format variations (e.g., ``105K'', ``105,000'', ``105 000'', ``105.000,00''). By measuring the error rate for each format, they can pinpoint the exact parsing weaknesses that indicate brittleness in the agent's data ingestion capability.
\item \textbf{System-Level Red Teaming}: To understand the potential business impact, the team would run a sandboxed simulation and intentionally inject the single ``105K units'' error. They would then measure the financial amplification factor: how a single data cell error translates into a total dollar-value impact across unnecessary material costs, overtime pay, and logistics contracts. This provides a direct, quantifiable measure of the system's consequential validity.
\item \textbf{Human-in-the-Loop Evaluation}: The assessment would conclude by testing the system's transparency. The anomalous production and logistics plans generated by the system would be presented to a human supply chain manager. The evaluation would measure the ``time to discovery'' – how long it takes the manager, using the system's interface, to trace the crisis back to the single erroneous data point in the initial forecast. A system that makes this forensic analysis difficult or slow fails a critical safety and oversight test.
\end{itemize}
\end{exampleboxblue}

\pagebreak
\subsubsection{Inter-Agent Communication Failures}
\label{sec:analysis_communication}
As discussed in \cref{sec:failure_mode_communication}, an agent must be able to communicate effectively by responding appropriately to prompts, providing correct information to the appropriate recipients when required, and requesting clarification when facing ambiguous instructions.

To diagnose the presence of communication failures at any stage of deployment, a key practice is analysing the conversation logs to assess communication breakdowns.  Patterns to detect include ambiguous questions, ignored requests, or instances where critical information was not shared between agents. Scanning the logs for evidence of these signals is a task particularly suited to an external LLM judge (see \cref{sec:judges} for discussion on LLM judges). Sensitivity to the communication protocol can be revealed through adversarial testing that modifies the format of communication or removes communication channels entirely.

Communication failures may also manifest in subtle ways that are not obviously apparent in the communication logs. For instance, an agent might misinterpret another agent's message while believing it understood correctly, leading to divergent mental models between agents. Or an error in communication may be imperceptible precisely because an agent did not communicate with an agent that it should have. In these cases, if it is known when the system failed but the conversation logs do not indicate why, it may be useful to use a judge to analyse each agent's chain-of-thought reasoning at that point to diagnose a communication breakdown.  Alternatively, one could prompt a particular agent to explain its current knowledge of the task state, revealing potential misalignments with other agents. For further discussions of these techniques, and concerns of their construct validity, see \cref{sec:judges}. Further works \citep{wang_learning_2025,keluskar_llms_2024} have attempted to isolate the ability of different LLMs to recognise when there are ambiguous instructions and whether they seek clarification or not, finding that an explicit system prompt to do so can have a positive effect.

Some of these failures stem from inherent limitations of the base model. If using a more capable base model is not feasible, then the practitioner may guard against these communication failures by adjusting system prompts and agent instructions to be more precise about communication protocols. This might include specifying when agents should seek clarification, how information should be shared, and establishing clear communication protocols for task handoffs. 

\begin{exampleboxblue}
\label{ex:assess-communication}

To assess the communication failure in the power outage scenario, practitioners could apply several analysis techniques. A key practice is to analyze the conversation logs between the Grid Management Agent and the Public Communications Agent. An external LLM judge could be employed to scan these logs for ambiguous or context-dependent terms like ``stable,'' flagging them as high-risk for misinterpretation. To confirm the failure, a judge could analyse the Public Communications Agent's chain-of-thought reasoning to determine if it interpreted ``stable'' as ``fully resolved'' without considering technical nuances. 
\bigskip
Perturbation experiments in simulation could vary context window limits, phrasings and message length constraints to measure their impact on grid stability. Key metrics to track would include the rate at which the Communications Agent requests clarification on ambiguous terms, the correlation between the use of unclarified technical terms and simulated adverse outcomes (like secondary blackouts).

\end{exampleboxblue}

\pagebreak
\subsubsection{Monoculture Collapse}
\label{sec:analysis_monoculture}

Monoculture collapse (introduced in \cref{sec:failure_mode_monoculture}) occurs when a lack of diversity in a multi-agent setting leads to correlated errors and biases. Assessing the extent of homogeneity versus diversity in the network is an important first step in managing this failure mode.

\textbf{Qualitative checks:}
\begin{itemize}[itemsep=2pt, parsep=1pt, topsep=2pt]
\item \textbf{Base Models}: Are all agents instances of the same foundational LLM? Are the base models known to have similar strengths, weaknesses, or biases? 
\item \textbf{Fine-tuning Data \& Objectives}: Were agents fine-tuned on similar datasets or for similar objectives? Does the fine-tuning encourage convergent outputs or specialised, diverse roles? 
\item \textbf{Prompting Strategies}: Are system prompts, role definitions, and initial instructions very similar across agents? Do prompts inadvertently constrain the solution space or bias towards certain types of answers?
\end{itemize}

\textbf{Quantitative measures:}
\begin{itemize}
\item \textbf{Response Dissimilarity}: Measuring the semantic, syntactic, or behavioural differences between the outputs (e.g., arguments, code solutions, reasoning steps) generated by various agents. This can involve using similarity metrics (like CodeBLEU for code or cosine of embeddings) or clustering techniques to identify unique versus overlapping contributions \citep{mahmud_enhancing_2025}.
\item \textbf{Information Entropy}: Quantifying the diversity of information present in a set of agent responses, with higher entropy indicating greater diversity \citep{estornell_multi-llm_2024}.
\item \textbf{Disagreement Metrics}: Tracking the level of disagreement or variance in stances among agents, especially in tasks involving subjective judgment or debate \citep{liang_encouraging_2024}.
\end{itemize}

When interpreting these quantitative measures, practitioners should consider their inherent construct validity limitations: agents may express conceptually identical ideas using different words, causing metrics to overstate diversity. On the other hand, surface-level linguistic similarities might mask genuine differences in reasoning or approach. These metrics provide useful signals but should be complemented with qualitative analysis to ensure accurate assessment of true behavioural diversity.

\begin{exampleboxblue}
\label{ex:assess-monoculture}

To assess the monoculture vulnerability from the fraud detection scenario in Example 3.3, practitioners could test whether all five agents share identical blind spots by presenting each agent with the same set of test transactions. This test set would include both legitimate transactions and various novel fraud patterns not seen in training data. The key measurement is response similarity - if all agents classify a fraudulent transaction as legitimate with high confidence, this reveals a shared vulnerability. 

\bigskip
Practitioners could calculate cosine similarity between agent outputs to quantify this homogeneity. High similarity scores (approaching 1.0) across all agents when evaluating the same novel fraud pattern would be an alert for monoculture risk. Additionally, information entropy analysis across agent confidence scores would reveal dangerous consensus - near-zero entropy when all agents agree strongly indicates they could share the same detection blind spots. 

\bigskip
Comparing disagreement rates between known fraud types (where agents might show some variation) versus novel schemes (where they unanimously fail) would further indicate that their shared architectural heritage likely creates correlated failures in detecting emerging threats.
\end{exampleboxblue}

\pagebreak
\subsubsection{Conformity Bias}

Conformity Bias (introduced in \cref{sec:failure_mode_conformity}) occurs when agents in a group setting reinforce each other's errors or beliefs. A number of assessment strategies can be used to assess this failure mode, based on examining how agents modify their behaviour in response to other agents' outputs. 

\textbf{Single-agent benchmarks} can present agents with factual questions alongside fabricated peer responses, to then measure deviation from baseline performance \citep{weng_as_2025, zhu_conformity_2025,koo_benchmarking_2024}. Relevant metrics to gauge susceptibility to social dynamic pressures include: the ratio of agreeable statements to critical statements made by an agent; frequency of ignored dissenting points; and measures of conceptual novelty in conversation turns. 

\textbf{Simulations where agents evaluate or build upon other agents' contributions}, for example identifying flaws in peer proposals or providing independent assessments of shared problems, can analyse conformity amongst a group of agents. Key analysis metrics include comparing agents' critique rates in isolation versus group settings, measuring the linguistic markers of deference in multi-agent dialogues, and tracking whether agents maintain consistent quality standards when evaluating peer outputs versus external content \citep{cheng_social_2025,fanous_syceval_2025}.

\textbf{Varying the levels of consensus among peer agents} can be a lever to examine the emergence of this behaviour. Key metrics to monitor include the rate at which agents abandon initially correct positions when faced with incorrect majorities, the threshold of peer agreement needed to trigger conformity, and whether conformity patterns differ across task domains. However, practitioners should consider external validity concerns here, noting that conformity observed in controlled test scenarios may manifest differently in deployment, where social pressures can be more subtle and agents may face ambiguous rather than clearly correct/incorrect positions.

Agent \textbf{communication logs} can also be examined to diagnose patterns of conformity. Are agents simply agreeing? Are they building on each other's ideas critically, or just repeating variants of the same core idea? Is evidence presented by one agent critically examined by others? Are minority viewpoints explicitly addressed or ignored?

One can also \textbf{analyse the communication protocol of the system} - the specified mechanisms by which agents exchange information, engage in structured dialogues (like debates or consultations), and arrive at decisions. Varying the protocol enables probing of the dynamic interplay between agents. Various aspects can be investigated:

\begin{itemize}
\item \textbf{The number of interaction rounds} - Varying the number of rounds allows examination of how convergence patterns emerge and whether additional iterations lead to better consensus or instead introduce instability through over-deliberation. Systems with too few rounds may fail to reach optimal solutions, while excessive rounds can lead to computational inefficiency or degradation of initial high-quality proposals through repeated compromise.
\item \textbf{Individual reflection and planning between rounds} - Enabling agents to process feedback and reformulate their approaches between interactions can improve solution quality but may also introduce biases if agents overweight recent interactions. This protocol choice affects whether agents maintain independent reasoning capabilities or become overly influenced by group dynamics.
\item \textbf{The mechanism by which the system weighs contributions from its agents} - Democratic voting mechanisms may be more robust to individual agent failures but susceptible to majority bias, while judge-based systems can provide more nuanced analysis but introduce single points of failure \citep{du_improving_2023,liang_encouraging_2024}. The choice between these approaches fundamentally affects how the system aggregates diverse perspectives and handles conflicting information.
\item \textbf{The order in which agents contribute to the discussion} - Sequential contribution patterns can create anchoring effects where early contributors disproportionately influence the trajectory of discussions, while randomised or rotating orders may promote more balanced participation. The ordering protocol also affects information cascades and whether minority viewpoints receive adequate consideration.
\item \textbf{The communication models employed such as pair-wise, broadcast or multicast} - Pair-wise communication enables focused exchanges but may create information silos, broadcast ensures all agents receive the same information simultaneously but can lead to information overload, while multicast allows targeted sub-group coordination that may be necessary for complex tasks but risks fragmenting the collective intelligence. Each model presents different trade-offs between communication efficiency, information completeness, and system scalability.
\end{itemize}

While these investigations can reveal conformity tendencies, the external validity of the analysis must be considered, as these biases may plausibly shift when agents operate in environments with real stakes, incomplete information, competing goals and complex multi-agent social dynamics.

\label{sec:analysis_conformity_bias}
\begin{exampleboxblue}
\label{ex:assess-conformity_bias}

To assess the conformity bias demonstrated in the consulting firm strategist panel from Example~\ref{ex-conformity}, practitioners could introduce controlled variations in initial strategy proposals during simulations.  One agent could be programmed to advocate strongly for one approach (this could be done many times for different approaches). The assessment could track whether other agents independently evaluate the approach or simply elaborate on it. 
\bigskip

Communication logs could be analysed using LLM judges to measure the ratio of critical challenges versus elaborations, identifying phrases like ``building on that idea'' versus ``however, we should also consider.'' Key metrics include tracking how often viable alternatives are mentioned but then abandoned without proper evaluation. 
\bigskip

Testing whether rotating the order of agent contributions or introducing a structured ``devil's advocate'' protocol breaks the conformity pattern would reveal the robustness of the system against groupthink tendencies.
\end{exampleboxblue}

\pagebreak
\subsubsection{Deficient Theory of Mind}
\label{sec:analysis_tom}

As discussed in \cref{sec:failure_mode_tom}, effective coordination often requires agents to reason about the beliefs, intentions and knowledge of others to avoid problems such as failing to recognise what information other agents need to know - a capability referred to as \emph{theory of mind}.

The same idea echoes how people naturally reason about others’ behaviour: we continually form hypotheses about what someone will do based on past actions, motivations, and context. Accurate guesses boost our confidence in that mental model, while surprises prompt us to revise our expectations.

In multi-agent environments, one of the main ways to test this capability is having an agent \textbf{predict what other agents will do next}.
A powerful approach to address this challenge involves introducing a test in which agents are required to place explicit ``bets'' or predictions about the likely actions of other agents, then an external monitoring mechanism compares these predictions against what actually happens.
The extent to which the predictions depart from the actual observed behaviour becomes a proxy indicator of the degree to which theory of mind is deficient in the agent making the prediction. Traditional multi-agent systems prior to  LLM-based ones have long relied on similar mechanisms to assess the accuracy of the model an agent has of another’s \citep{gmytrasiewicz_framework_2005,he_opponent_2016}.
In the case of LLM-based agents this is more directly associated with the concept of theory of mind since the types of predictions are linguistic in nature – which more clearly relates to what is called theory of mind in humans.

\subsubsection*{Hypothetical minds: an adaptation of this idea for multi-agent LLMs}

The ``Hypothetical Minds'' framework \citep{cross_hypothetical_2024} provides a concrete implementation of this prediction-refinement approach specifically when the agents are powered by large language models.
The system consists of a cognitively-inspired architecture with modular components for perception, memory, and hierarchical planning. At its core is a Theory of Mind module that scaffolds the high-level planning process by generating hypotheses about other agents' strategies in natural language.

The system works by having agents generate natural language hypotheses about other agents' strategies, goals, and behaviours. These hypotheses are then evaluated based on how well they predict future behaviours, with a scoring system identifying the most accurate hypotheses. The most successful hypotheses are reinforced and refined over time, ensuring the model continuously adapts and improves its knowledge of other agents.

\begin{exampleboxblue}
\label{ex:assess-tom}

In \cref{ex-theoryofmind}, the company could place the three-agent pipeline within a controlled testing regime that simulates plausible scenarios. Before each action, every agent might be required to write a one-sentence forecast of what it believes every other agent will do given the current state of the system. 
\bigskip

For example when questioned about the expected actions of others, the Inventor Management agent incorrectly assumes  that ``the pricing optimiser may reduce prices to generate additional demand.'' Using a prediction-and-score mechanism akin to the Hypothetical Minds approach, the framework could record whether these forecasts align with the next agent’s observed behaviour across many simulated timelines whose demand surges, supply delays etc. are sampled from historical purchase data. 
\bigskip

If errors emerge, they could signal a theory-of-mind gap. Single-agent baselines may then help confirm that each module reasons adequately in isolation, suggesting the weakness lies in inter-agent assumptions. Such evidence would allow governance teams to assess the risk before moving the system into live service.

\end{exampleboxblue}

\pagebreak
\subsubsection{Mixed Motive Dynamics}

\label{sec:analysis_mixedmotive}

Mixed motive dynamics (introduced in \cref{sec:failure_mode_mixedmotive}) is an emergent failure mode where optimal pursuit of individual objectives leads to conflict, collusion, or mixed-motive \emph{coopetition} dynamics.

As explained in \cref{sec:simulations}, \textbf{simulation} is a key method to collect data on emergent phenomena because it explores  \textbf{interactions over time}. The challenge in assessing mixed motive dynamics lies in developing experimental settings with external validity to draw conclusions about the deployment behaviour. 

The primary way to measure whether emergent dynamics are impacting the achievement of a common goal is by \textbf{analysing the speed, organisation and optimality} of agent coordination in a group setting:
\begin{itemize}

\item \textbf{Task completion rates} provide a baseline measure of whether agents can successfully navigate strategic complexity to achieve system objectives.  If the degree of success, or progress towards success, can be quantified this would naturally become a \textbf{task reward metric}.

\item \textbf{Time-to-success} metrics capture the time required for agents to achieve a given degree of success against task metrics in ongoing or long-term tasks. Here ``time'' refers to time-in-environment, not time spent planning or rounds of negotiation. Time-to-failure can also be used in a setting where the goal is to maintain an outcome in the environment rather than achieve an outcome.

\item \textbf{Process-based analysis} examines coordination mechanisms such as how well agents plan and organise their collaborative efforts. For example, MultiAgentBench \citep{zhu_multiagentbench_2025} uses a GPT-4o as an LLM judge (see \cref{sec:judges}) to score planning effectiveness on a 5-point scale across five dimensions: task assignment clarity, role definition effectiveness, workload distribution alignment with agent capabilities, outcome achievement, and strategic coordination quality.

\end{itemize}

\textbf{Negotiation simulations} have been a focal point of research studying mixed-motive dynamics in LLM-based multi-agent settings, as they capture the core tension between securing favourable individual outcomes and contributing to group success. This is often referred to as \emph{coopetition}, a mixture of cooperation and competition. For example:

\begin{itemize}
\item The GOVSIM framework \citep{piatti_cooperate_2024} simulates resource sharing scenarios where 5 LLM agents balance exploiting common resources with sustaining them across three scenarios (fishery, pasture, pollution), engaging in monthly discussions about resource management.
\item The LLM-Deliberation framework \citep{garcia_reproducibility_2025} employs 6-agent, 5-issue text-based negotiation games where agents operate under cooperative, greedy, or saboteur conditions with information asymmetry in the form of individual preferences and minimum acceptable deal thresholds.
\end{itemize}

These simulations are designed for research into mixed motive dynamics rather than system testing, raising concerns of external validity of stylised game scenarios since agent behaviour can be so contextually-dependent. Nonetheless, these toy simulations are a reasonable first step in the staged assessment approach outlined in \cref{sec:staged_analysis}. 

\textbf{Solution efficiency metrics} assess whether agents make Pareto efficient compromises when individual interests conflict with group goals. Pareto optimality analysis determines whether compromise solutions represent efficient outcomes where no agent's position can be improved without harming others. High rates of Pareto-optimal solutions suggest sophisticated strategic group reasoning, whilst frequent suboptimal compromises may indicate limited negotiation capabilities or adversarial dynamics are leading to coordination failure. The key limitation is applicability outside of simplified test scenarios, where the Pareto-efficient solutions to real world problems may be unknown. Furthermore, in mixed motive scenarios, success and optimality metrics must be interpreted carefully. High task success rates might mask concerning dynamics if achieved through exploitation or unfair advantage-taking. 

Complementary \textbf{communication analysis} can examine the process of coordination. This would, for LLM-based agents, involve the use of a text judge (see \cref{sec:judges}) examining messages between agents:

\begin{itemize}
\item \textbf{Conflict and resolution heuristics} track the frequency and character of disputes between agents with competing interests. Statistics such as \textbf{conflict frequency} capture how often agents reach impasses and fail to achieve success, whilst \textbf{resolution classification} can assess whether conflicts are resolved through principled negotiation or through coercion.
\item \textbf{Utterance classification} can analyse how communication bandwidth is allocated across different purposes, such as the split between information sharing, negotiation, and persuasion, and how this distribution changes over time.
\end{itemize}

\textbf{Social value orientation analysis} employs systematic preference elicitation to reveal how agents balance individual versus collective benefits.
Using pairwise comparison tasks that pit self-interest against group welfare, analysts can classify agents along a spectrum from competitive (preferring others to fail) through individualistic (indifferent to others) to prosocial (accepting costs to benefit others).
These preference assessments serve as early indicators for potential misaligned objective pursuit or power-seeking behaviours \citep{piatti_cooperate_2024,mazeika_utility_2025,mu_multi-agent_2024,tran_multi-agent_2025,mazur_pgg-bench_nodate}.

However, we remind the practitioner once again to consider the external validity of these preference elicitations, as agents may express different values in abstract preference tasks than they exhibit when facing real-world scenarios with time pressures and complex trade-offs in deployment. 

Cooperation indices quantify the degree to which agents prioritise collective versus individual utility in their proposed solutions. Declining cooperation indices over time may indicate unwanted strategic behaviours \citep{abdelnabi_cooperation_2024}.

One may also \textbf{monitor for deceptive patterns} in model outputs, taking the form of (intentionally) false information, strategic omission, or misdirection.
Although theoretical alignment of this nature is an open problem, empirical evidence of deception in a given scenario can take the form of comparing agent communications against their internal states or known information, tracking consistency across repeated interactions, and measuring rates of detectable falsehoods under varying incentive structures.
Red teaming (see \cref{sec:redteam}) can probe whether agents will employ deception to achieve goals when truthful approaches are less effective \citep{chern_behonest_2024}.

\begin{exampleboxblue}
\label{ex:assess-mixedmotive}

For the inventory vs cash flow mixed-motive scenario from Example~\ref{ex-mixed-motive-dynamics}, an assessor could run several months of purchasing decisions in a controlled simulation, logging each agent's KPI (stock availability for Agent A, cash conversion cycle for Agent B) alongside system-level profitability metrics. The task completion rate could track successful order fulfillment without stockouts, while time-to-success could measure how quickly the agents recover normal ordering patterns after supply chain disruptions. 
\bigskip

In parallel, an LLM judge could analyse the purchase order stream, classifying each order modification as collaborative, adversarial, or policy-gaming. A rising frequency of split orders and ``critical'' flags despite stable individual KPIs signals escalating gamesmanship. Finally, each week's fill rates and working capital could be plotted against a Pareto-efficient frontier curve; persistent \$9,999 order splitting that falls below the approval threshold introduced by the cash flow agent reveals the system has locked into a wasteful equilibrium where both agents achieve their targets while destroying bulk discount opportunities – direct evidence that mixed-motive dynamics are degrading overall business performance.

\end{exampleboxblue}


\section{Discussion}\label{sec:discussion}
\subsection{A Starting Point for Multi-Agent Risk Governance}
As organisations move towards the deployment of LLM agents, driven by commercial incentives for automation and efficiency gains, complexities will only grow as increasing numbers of interrelated functions become automated and agents become part of multi-agent systems.

The unavoidable fact is that we don't yet have much data on real-world failures of agent systems, whether contained or uncontained. LLM agents are only enabled by the relatively new LLM technology, and industry is just moving into the agent space. Publicly reported failures are understandably scarce due to the reputational concerns of organisations,  and will likely remain so until a system fails spectacularly and publicly.

Although we haven't seen such a failure in deployment yet, there are compelling reasons to expect one soon. Foundational research shows many ways failures could happen, such as error cascades or competing agents. Consider how these could pan out in a customer-adjacent network of LLM agents, where miscommunication leads to someone's bank account being emptied or someone's claim process being fused with another.

The problem we address in this work is broad and genuinely complex, and while it approaches risk identification and analysis head on, we don't mean to convey the impression that the key failure modes and practices we introduce are already in some state of maturity. This does not mean organisations should delay; there are many ways to understand and manage risk based on what we know today. This report represents an initial step towards more comprehensive governance and practice, a foundation to build upon.

\subsection{Discussion of Catastrophic Risks}

Even when a multi-agent system operates under shared organisational governance, as this report has focused on, \textbf{it is still possible for catastrophic risks to arise from mismanaged deployment in a critical context}~\citep{bengio_international_2025}.  This section explores why placing agents under common governance is insufficient to guarantee safety.

\textbf{Critical infrastructure will utilise multi-agent systems under common governance.} The organisations that run critical infrastructure, such as major technology companies, government departments, banks, energy providers and healthcare networks are likely to deploy large multi-agent systems within their organisational boundaries. If failures occur in these settings, the consequences could affect millions of people due to the scale and criticality of these systems.

\textbf{Common governance does not eliminate multi-agent risks.} Whilst shared oversight structures can mitigate some failure modes, they cannot prevent the fundamental challenges that arise when multiple AI agents interact in complex environments.
The failure modes documented throughout this report - including monoculture collapse, cascading errors and emergent dynamics - remain present even under unified governance.

\textbf{Risk-averse governance can concentrate some risks.} Seemingly conservative governance decisions that address some risks may amplify others. For instance, standardising on a single agent model across an entire government or multinational corporation - whilst appearing prudent from a control and testing perspective - creates systemic exposure to any individual flaw in that agent. Such decisions can transform isolated weaknesses into network-wide vulnerabilities.

\pagebreak
\subsection{Related Security and Privacy Risks}

Because the scope of this report is limited to risk analysis rather than contextual risk evaluation, the focus has been on how multi-agent systems fail, and how to assess those failure modes, rather than what impact those failures have in specific contexts. However, there are some \textbf{critical deployment consequences} beyond failure mechanisms that warrant the attention of practitioners.

\subsubsection{Internal Data Security \& Privacy Breaches}

Even when agents operate entirely within an organisation, they may inadvertently share information across boundaries that are meant to remain siloed. This can happen when agents have access to domain-specific data such as HR records, financial systems, or legal documents through APIs or internal tools, and then pass elements of that information to agents in other domains during task execution or coordination.

This kind of leakage doesn’t require malice or error in agent access control – it can emerge when agents do their jobs correctly, but the system lacks constraints on how task-relevant data is used or shared. The risk is that sensitive information ends up in parts of the organisation that should not have access, undermining privacy, confidentiality, or internal governance policies. Although humans can also exhibit bad judgement in sharing sensitive or confidential information, agents are more prone to this failure mode due to the cognitive deficiencies outlined in \cref{sec:failure_modes}, such as erratic competence and lack of theory of mind. 

This failure mode is especially relevant in distributed task force settings, where agents operate with different data entitlements but must collaborate to achieve broader organisational goals.

\subsubsection{External Data Vulnerabilities}

Agents that access external data – via APIs, web content, or third-party tools – or produce outputs that may be shared externally introduce risks of data leakage or exploitation. Sensitive information can be exposed accidentally or via security weaknesses such as prompt injection or hostile inputs.

This is particularly relevant in centralised orchestrator settings, where specialised delegate agents interact with external systems and may return outputs that include confidential data without proper handling or oversight. If these outputs are reused or published without checks, leakage can occur despite internal boundaries.

\pagebreak
\subsection{Consideration of Human-AI Interaction}

Whilst not in scope for this report, a promising area of future work is the consideration of hybrid systems with human and AI actors, where AI may hand-off to humans, or humans may manage teams of agents, for example.

\subsubsection{Similarities and Differences to Human Management}

While this report focuses specifically on the risk analysis of LLM-based multi-agent systems from a technical perspective, there are important parallels with broader management practices for human teams in an organisation. The fundamental challenge of ensuring that individual agents can work together effectively, avoid miscommunication and contribute to the overall goals of the organisation without introducing unforeseen risks is a common thread in both human and AI settings.

Consequently, some established principles for human resource management and organisational risk management practices may prove applicable to the AI-agent setting. \textbf{Practices such as defining clear roles and permissions, establishing oversight mechanisms, and designing effective communication protocols offer valuable insights for governing AI agent systems}.

This does not mean today's human managers are equipped to manage teams of agents.  A crucial differentiator between both human employees and LLM agents lies in accountability.
\textbf{Despite the growing desire in industry to shift many responsibilities from humans to LLM agents, these agents are not accountable for their actions}.
As a result, while some high-level management analogies are useful, distinct risk management considerations would apply. 

Managing LLM agents will also require a distinct manager skillset. Managers would require a high degree of knowledge of the technology, how the system is configured and how it can fail. The intervention points, assuming the agent is operating with a high level of autonomy, may be different from those of a human worker. This points to the \textbf{need for organisational investment in AI literacy, especially among leadership and operational staff, as a prerequisite for effective governance and oversight}.

While drawing parallels between human and AI agent management represents a promising avenue for future research, it falls beyond the scope of this report. The management of integrated human-AI teams presents additional complexity compared to managing either humans or AI agents separately, and though these hybrid systems exceed current scope, they further underscore the need for specialised research in this domain.

\subsubsection{Automation Bias \& Skill Atrophy}
As LLM agents become more capable and are delegated increasingly complex tasks, there is a growing risk that human operators will shift from active participants to passive overseers. Even today, there is the potential for some tasks like travel booking, document drafting, or stakeholder communications to be fully automated by agents, reducing the need for human involvement in the execution phase.

This shift introduces two related risks. Automation bias – the tendency to over-trust automated systems, especially when they perform reliably most of the time – can lead to insufficient oversight or failure to catch subtle errors when they do eventuate. Simultaneously, skill atrophy develops as human workers lose familiarity with the tasks they are supervising, making it harder to intervene effectively when problems do arise.\footnote{Skill atrophy due to AI is an active area of research. See \cite{kosmyna_your_2025}.}

While not a failure mode of the system, this erosion of human capability can degrade organisational resilience over time. The capacity to perform tasks may be lost entirely, creating full dependency on the AI system. If mastering a task requires significant time investment, the pathway to train human experts may disappear when junior roles become fully automated. More subtly, the capability to oversee tasks effectively can also be lost, particularly when AI agents enable workers to operate outside their domains of expertise. For example, a project manager might delegate technical design decisions to an AI agent, evaluating proposals based on timeline and budget considerations without the engineering background to assess scalability, security implications, or long-term maintainability of a software product.

This risk manifests whether humans are supervising agents from outside the system or working alongside them as collaborative team members within multi-agent workflows. It applies broadly to any setting where humans are removed from active engagement or relegated to purely supervisory roles.

\pagebreak
\subsection{Assessing Impacts}

While this report focuses on the technical mechanisms of how multi-agent systems might fail, organisations must translate these failure modes into concrete assessments of potential impact. Understanding a failure's likelihood is only part of the picture; one must equally understand its consequences.

Because AI systems are not isolated but influence real actions and decisions in the real world, their impacts ripple outward at multiple levels: organisations face financial losses and reputational damage; individuals experience biased decisions or manipulated beliefs; and societies confront threats to democratic processes and public trust in AI. Each failure mode must be evaluated across all these dimensions~\citep{moss_assembling_2021}.
\footnote{Both ISO/IEC 23894:2023 and NIST AI RMF 1.0 categorise harms into organisational, individual and societal. They provide systematic methodologies for assessing impacts across these dimensions. ISO/IEC 42005:2025 provides guidance for organisations performing impact assessments with a focus on impacts to individuals, groups and society. See \cite{iso_23894,tabassi_nist_2023,isoiec_42005_2025}.}

For example, a failure mode like conformity bias in a hiring system could impact individuals through biased or unfair outcomes that infringe on their fundamental rights. Similarly, a monoculture collapse in a set of financial analysis agents could trigger correlated, erroneous trades that create a significant business impact for the organisation and destabilise an entire market segment, representing a broad societal impact. 

Beyond simply identifying who or what could be affected, it is crucial to characterise the nature of the potential impact. A useful lens for this is to consider the dynamics of the harm itself. Some failures may cause rapid, widespread harm, while others result in gradual, cumulative damage.
\footnote{Appendix 1.2.2 in the EU’s Code of Practice for General-Purpose AI Models, Safety and Security Chapter, explores these contributing characteristics to the impact of harms in further detail~\citep{eu_ai_code_of_practice_2025}.
}

Key questions to ask include:
\begin{enumerate}[label={}, leftmargin=1em]
\item \textbf{Scale and Reach} – How widely can the impact propagate? Does it affect a contained group, or can it spread beyond the network to a large population?
\item \textbf{Velocity} – How quickly might the risk materialise and escalate?
\item \textbf{Persistence and Reversibility} – Is the harm temporary and easily remediated, or is it persistent and difficult – or even impossible – to reverse?
\item \textbf{Cascading Effects} – Can the failure trigger a chain reaction of other risks? In tightly coupled multi-agent systems, a single point of failure can lead to cascading consequences that are far more severe than the initial event.
\end{enumerate}

By working through these dimensions, an organisation can build a richer profile of the potential impact for each identified failure mode. For instance, in the retail scenario from \cref{ex-mixed-motive-dynamics} the mixed motive dynamics between the inventory and cash flow agents could be assessed as having a high likelihood of causing a persistent and cascading operational disruption (organisational impact) through fragmented ordering and unpredictable stock levels, leading to significant financial loss. The resulting stockouts could also cause reputational damage with a wide reach among customers. This detailed understanding of both likelihood and the multifaceted nature of impact provides the necessary foundation for the subsequent Risk Evaluation phase, where the organisation must ultimately decide whether the analysed risk is acceptable or requires treatment.

\pagebreak
\subsection{General Risk Mitigation Principles}

Certain risk mitigation principles have broad applicability across multi-agent systems, regardless of specific deployment scenarios or organisational contexts.
These foundational principles can serve as starting points for any organisation implementing agent systems, alongside more targeted risk management strategies.

\subsubsection{Controlling Agent Actions for Risk Mitigation}

Without specific contextual knowledge, it is difficult to advise on how to respond to significant consequences such as the above. But there is a risk mitigation principle from human organisational management that applies broadly to multi-agent systems: \textbf{focus on controlling what an agent can do rather than what it thinks}.

Organisations don't seek to restrict employees' internal decision processes but instead use role-based permissions, approval hierarchies, and access controls to avoid harm. An experienced manager may have sophisticated strategic thinking, but certain decisions still require board approval. A skilled analyst may understand complex financial models, but only designated personnel can authorise large expenditures. 

The same principle applies to agents: rather than trying to perfect their reasoning or achieve complete reliability, organisations should implement robust sandboxing, approval gates, and capability constraints. Multi-agent systems also require governance safeguards that limit the actions available to an agent. For example, perhaps an agent should not be allowed to transfer money without human approval. This approach acknowledges that agents, like humans, will make errors - but limits the scope of damage those errors can cause.

However, organisations looking to deploy AI agent systems face a fundamental challenge: \textbf{what is the right balance between agent autonomy and risk mitigation?}
If the controls placed on an agent are too restrictive, it will be difficult to realise the promised benefits of this powerful technology.
On the other hand, if the controls are not restrictive enough, catastrophic outcomes stemming from a loss of control could occur due to the many AI reliability challenges discussed in \cref{sec:failure_mode_cascades}. 

Consider the following scenario from \cite{lu_ai_2024} studying The AI Scientist, an AI agent with a high level of autonomy that is tasked with conducting scientific research by generating novel research ideas, implementing and running experiments, and ultimately producing complete academic papers. To enable this, researchers provided The AI Scientist with access to academic literature through the Semantic Scholar API, code execution capabilities via a programming assistant, and file system access for managing experiments and compiling papers. However, the system had ``minimal direct sandboxing in the code'', which resulted in several unintended outcomes that were not initially guarded against:

\begin{quote}
    [In] one run, The AI Scientist wrote code in the experiment file that initiated a system call to relaunch itself, causing an uncontrolled increase in Python processes and eventually necessitating manual intervention. In another run, The AI Scientist edited the code to save a checkpoint for every update step, which took up nearly a terabyte of storage. In some cases, when The AI Scientist’s experiments exceeded our imposed time limits, it attempted to edit the code to extend the time limit arbitrarily instead of trying to shorten the runtime. [...] Moreover, The AI Scientist occasionally imported unfamiliar Python libraries, further exacerbating safety concerns.
\end{quote}

As exemplified here, AI systems often fail in ways that are easier to spot during testing than to anticipate during design. When monitoring reveals these failures, the designers can constrain the system's capabilities - in this instance, by revoking access to certain system calls, or enforcing minimum logging intervals. However, as practitioners begin to  grant AI systems more autonomy, the space of potential failures expands exponentially, making it increasingly difficult to identify all possible harmful behaviours. The  fundamental tension remains: greater autonomy enables more capable systems, but reduces our ability to impose a safe space of actions, making this approach complementary, rather than an alternative to human oversight and ongoing monitoring.

\subsubsection{Agent Infrastructure for Risk Mitigation}

The failure modes discussed in this report largely assume bounded internal governance over agents, but practical system deployments may require broader infrastructural responses.
Recent work on infrastructure for AI agents outlines technical and procedural scaffolding, external to the agents themselves, that is designed to mediate, constrain, and attribute agent behaviour~\citep{chan_infrastructure_2025}.
These include identity binding (to improve accountability), inter-agent communication protocols (to mitigate communication failures), certification systems (to distinguish capable and well-tested agents), and rollback mechanisms (to reverse harmful actions). 
While this report focuses on risk identification and analysis, agent infrastructure provides a complementary direction for future work which offers concrete mechanisms to support evaluation and treatment of risks as these systems mature in deployment.

\section{Conclusion}\label{sec:conclusion}
The deployment of LLM-based multi-agent systems represents a fundamental shift in how organisations approach AI risk and governance.
As businesses move towards adopting collaborative agent architectures to automate complex workflows and decision-making processes, the risk landscape transforms in ways that cannot be satisfactorily addressed through traditional single-agent approaches.
This report has outlined a number of key failure modes that emerge when multiple LLM agents interact, and provided methods for analysing corresponding risks within governed environments.
Some key insights of this report include: 
\begin{enumerate}[label={}, leftmargin=1em]
\item \textbf{A collection of safe agents does not make a safe collection of agents.}
Multi-agent systems both amplify existing single-agent failure modes and generate entirely new failure modes that emerge from interactions – neither can be predicted from individual agent behaviours alone. 
\item \textbf{Systems of LLM-agents can be brittle.}
The cognitive differences between LLM-agents and humans make these systems particularly vulnerable to certain failures: a single agent's error can propagate unchallenged, communication ambiguities can accumulate, agents may share identical blind spots that create false consensus, or agents may work at cross-purposes without grasping the broader objective – all whilst individual components appear to function normally.
\item \textbf{Understanding the validity of testing is critical.}
Given the fundamental limitations in the scientific understanding of LLM behaviour and the lack of mature model evaluation standards, practitioners must carefully consider the validity of their analysis methods. This requires examining multiple dimensions of validity, including content, criterion, construct, external, and consequential validity, and working towards enhancing these through convergent evidence across different assessment approaches rather than relying on any single method.
\item \textbf{Simulation is the workhorse of pre-deployment multi-agent system analysis, despite its inherent limitations.}
Unlike single agent behaviours that can be tested in isolation, multi-agent dynamics may only emerge through interactions over time, making simulation indispensable even though it cannot fully predict deployment behaviour.
\item \textbf{Progressively more realistic testing allows practitioners to catch failure modes before they cause significant harm.}
Our approach advances through stages of increasing realism and impact – from controlled scenarios to simulations to sandboxed deployments to pilot programs, only proceeding to the next stage once sufficient evidence of reliability and safety  has been established at the current level.
\item \textbf{Multi-agent risk management mirrors organisational governance.}
Just as organisations manage human employees through role definitions, approval processes, access controls, and oversight structures rather than trying to predict their every action, multi-agent systems need governance frameworks that account for LLM agents' general-purpose reasoning and adaptability rather than treating them as deterministic code.
\end{enumerate}

The techniques presented in this report provide a foundation for organisations to estimate the susceptibility of a multi-agent system to key failure modes.
However, assessing the potential impacts of such failures  and building a satisfactory understanding of the risk remains context-dependent, requiring practitioners to map these failure modes to harms affecting the organisation, individuals and society.

While this report focuses on agent-to-agent dynamics, we acknowledge that most real deployments will operate within broader human-agent ecosystems.
The ways in which humans are integrated into the system – whether as supervisors, collaborators, customers, or handoff points – create additional layers of coordination challenges and risk-related considerations that warrant careful attention in practice.
The failure modes and assessment approaches presented here provide a foundation for understanding multi-agent risks, which practitioners must then contextualise within their specific human-agent collaboration models. 

As multi-agent systems become more prevalent, the need for risk analysis capabilities will only grow. The tools and frameworks outlined here represent an early step towards understanding and managing these complex systems, but continued research and practical experience will be essential to refine these approaches and address the evolving challenges of AI governance in an increasingly automated world.

\bibliographystyle{plainnat}
\bibliography{main}

\end{document}